\documentclass[a4paper,conference]{IEEEtran}
\usepackage{graphicx}
\usepackage{algorithm,algorithmic}
\usepackage{setspace}
\usepackage[font=scriptsize]{subcaption}
\usepackage[font=footnotesize]{caption}
\usepackage{mathtools}
\usepackage{amsthm}
\usepackage[draft]{hyperref} 
\usepackage{cite}
\usepackage{amssymb}
\usepackage{physics}

\usepackage{tikz}
\usepackage{pgfplots}
\pgfplotsset{compat=newest}
\pgfplotsset{plot coordinates/math parser=false}
\usetikzlibrary{plotmarks,shapes,patterns,decorations.pathreplacing,backgrounds,calc,arrows,arrows.meta,spy,matrix,shadows,trees,positioning}
\usepgfplotslibrary{patchplots,groupplots}
\usepackage{tikzscale}
\usepackage{tikz-qtree}

\usepackage[nowarn,acronyms,nonumberlist,nopostdot,nomain,nogroupskip]{glossaries}
\usepackage{xcolor}

\makeatletter
\newcommand{\LeftEqNo}{\let\veqno\@@leqno}
\makeatother

\makeatletter
\newcommand\fs@spaceruled{\def\@fs@cfont{\bfseries}\let\@fs@capt\floatc@ruled
  \def\@fs@pre{\vspace{0.085cm}\hrule height.8pt depth0pt \kern2pt}%
  \def\@fs@post{\kern2pt\hrule\relax}%
  \def\@fs@mid{\kern2pt\hrule\kern2pt}%
  \let\@fs@iftopcapt\iftrue}
\makeatother

\newacronym{3gpp}{3GPP}{3rd Generation Partnership Project}
\newacronym{5g}{5G}{5\textsuperscript{th} Generation}
\newacronym{5gc}{5GC}{5G Core}
\newacronym{bs}{BS}{Base Station}
\newacronym{abft}{A-BFT}{Association-BeamForming Training}
\newacronym[firstplural=Access Categories (ACs)]{ac}{AC}{Access Category}
\newacronym{adc}{ADC}{Analog to Digital Converter}
\newacronym{addts}{ADDTS}{Add Traffic Stream}
\newacronym{afbw}{AFBW}{Average Fading Bandwidth}
\newacronym{aid}{AID}{Association ID}
\newacronym{aimd}{AIMD}{Additive Increase Multiplicative Decrease}
\newacronym{am}{AM}{Acknowledged Mode}
\newacronym{amc}{AMC}{Adaptive Modulation and Coding}
\newacronym{ampdu}{A-MPDU}{MAC Protocol Data Unit Aggregation}
\newacronym{aoa}{AoA}{Angle of Arrival}
\newacronym{aod}{AoD}{Angle of Departure}
\newacronym{ap}{AP}{Access Point}
\newacronym{app}{APP}{Application Layer}
\newacronym{aqm}{AQM}{Active Queue Management}
\newacronym{ar}{AR}{Augmented Reality}
\newacronym{arf}{ARF}{Auto Rate Fallback}
\newacronym{arp}{ARP}{Address Resolution Protocol}
\newacronym{ati}{ATI}{Announcement Transmission Interval}
\newacronym{awgn}{AGWN}{Additive White Gaussian Noise}
\newacronym{awv}{AWV}{Antenna Weight Vector}
\newacronym{balia}{BALIA}{Balanced Link Adaptation}
\newacronym{bdp}{BDP}{Bandwidth-Delay Product}
\newacronym{ber}{BER}{Bit Error Rate}
\newacronym{bf}{BF}{Beamforming}
\newacronym{bframe}{B-frame}{Bipredictive-coded frame}
\newacronym{bhi}{BHI}{Beacon Header Interval}
\newacronym{bi}{BI}{Beacon Interval}
\newacronym{brp}{BRP}{Beam Refinement Protocol}
\newacronym{bss}{BSS}{Basic Service Set}
\newacronym{bti}{BTI}{Beacon Transmission Interval}
\newacronym{cad}{CAD}{Computer-aided Design}
\newacronym{cbap}{CBAP}{Contention-Based Access Period}
\newacronym{cbr}{CBR}{Constant Bitrate}
\newacronym{cc}{CC}{Congestion Control}
\newacronym{cdf}{CDF}{Cumulative Distribution Function}
\newacronym{cir}{CIR}{Channel Impulse Response}
\newacronym{cn}{CN}{Core Network}
\newacronym{cp}{CP}{Control Plane}
\newacronym{cqi}{CQI}{Channel Quality Indicator}
\newacronym{crs}{CRS}{Cell Reference Signal}
\newacronym{csirs}{CSI-RS}{Channel State Information - Reference Signal}
\newacronym{csmaca}{CSMA/CA}{Carrier Sense Multiple Access with Collision Avoidance}
\newacronym{cts}{CTS}{Clear to Send}
\newacronym{dc}{DC}{Dual Connectivity}
\newacronym{dce}{DCE}{Direct Code Execution}
\newacronym{dcf}{DCF}{Distributed Coordination Function}
\newacronym{dci}{DCI}{Downlink Control Information}
\newacronym{delts}{DELTS}{Delete Traffic Stream}
\newacronym{dl}{DL}{Downlink}
\newacronym{dmg}{DMG}{Directional Multi-Gigabit}
\newacronym{dmr}{DMR}{Deadline Miss Ratio}
\newacronym{dmrs}{DMRS}{DeModulation Reference Signal}
\newacronym{dti}{DTI}{Data Transmission Interval}
\newacronym{e2e}{E2E}{End-to-End}
\newacronym{ecn}{ECN}{Explicit Congestion Notification}
\newacronym{edca}{EDCA}{Enhanced Distributed Channel Access}
\newacronym{edf}{EDF}{Earliest Deadline First}
\newacronym{embb}{eMBB}{Enhanced Mobile BroadBand}
\newacronym{enb}{eNB}{evolved Node Base}
\newacronym{endc}{EN-DC}{E-UTRAN-\gls{nr} \gls{dc}}
\newacronym{epc}{EPC}{Evolved Packet Core}
\newacronym{es}{ES}{Edge Server}
\newacronym{ese}{ESE}{Extended Schedule Element}
\newacronym{fdd}{FDD}{Frequency Division Duplexing}
\newacronym{fdma}{FDMA}{Frequency Division Multiple Access}
\newacronym{fec}{FEC}{Forward Error Correction}
\newacronym{fov}{FoV}{Field-of-View}
\newacronym{fs}{FS}{Fast Switching}
\newacronym{ftp}{FTP}{File Transfer Protocol}
\newacronym{fr2}{FR2}{Frequency Range 2}
\newacronym{gmm}{GMM}{Gaussian Mixture Model}
\newacronym{gnb}{gNB}{Next Generation Node Base}
\newacronym[firstplural=Group of Pictures (GoPs)]{gop}{GoP}{Group of Pictures}
\newacronym{harq}{HARQ}{Hybrid Automatic Repeat reQuest}
\newacronym{hetnet}{HetNet}{Heterogeneous Network}
\newacronym{hh}{HH}{Hard Handover}
\newacronym{hol}{HOL}{Head-of-Line}
\newacronym{hqf}{HQF}{Highest-quality-first}
\newacronym{ia}{IA}{Initial Access}
\newacronym{iab}{IAB}{Integrated Access and Backhaul}
\newacronym{ibss}{IBSS}{Independent Basic Service Set}
\newacronym{id}{ID}{Identifier}
\newacronym{ifi}{IFI}{Inter-Frame Inter-arrival}
\newacronym{iframe}{I-frame}{Intra-coded frame}
\newacronym{imt}{IMT}{International Mobile Telecommunication}
\newacronym{imt2020}{IMT-2020}{International Mobile Te\-le\-com\-mu\-ni\-ca\-tion-2020}
\newacronym{inr}{INR}{Interference to Noise Ratio}
\newacronym{iot}{IoT}{Internet of Things}
\newacronym{ipa}{IPA}{Inter-Packet Arrival}
\newacronym{ism}{ISM}{Industrial, Scientific, and Medical}
\newacronym{itu}{ITU}{International Telecommunication Union}
\newacronym{kpi}{KPI}{Key Performance Indicator}
\newacronym{lcf}{LCF}{Level Crossing Frequency}
\newacronym{lcm}{lcm}{least common multiple}
\newacronym{lcr}{LCR}{Level Crossing Rate}
\newacronym{los}{LoS}{Line-of-Sight}
\newacronym{lp}{LP}{Low Power}
\newacronym{lte}{LTE}{Long Term Evolution}
\newacronym{m2m}{M2M}{Machine to Machine}
\newacronym{mac}{MAC}{Medium Access Control}
\newacronym{mc}{MC}{Multi-Connectivity}
\newacronym{mcs}{MCS}{Modulation and Coding Scheme}
\newacronym{mec}{MEC}{Mobile Edge Cloud}
\newacronym{mi}{MI}{Mutual Information}
\newacronym{mib}{MIB}{Master Information Block}
\newacronym{mimo}{MIMO}{Multiple Input, Multiple Output}
\newacronym{mumimo}{MU-MIMO}{Multi-User Multiple Input, Multiple Output}
\newacronym{ml}{ML}{Machine Learning}
\newacronym{mlr}{MLR}{Maximum-local-rate}
\newacronym[plural=\gls{mme}s,firstplural=Mobility Management Entities (MMEs)]{mme}{MME}{Mobility Management Entity}
\newacronym{mmw}{mmW}{Millimeter Wave}
\newacronym{moi}{MoI}{Method of Images}
\newacronym{mpc}{MPC}{Multi Path Component}
\newacronym{mpdu}{MPDU}{MAC Protocol Data Unit}
\newacronym{mptcp}{MPTCP}{Multipath TCP}
\newacronym{mr}{MR}{Maximum Rate}
\newacronym{mrdc}{MR-DC}{Multi \gls{rat} \gls{dc}}
\newacronym{msdu}{MSDU}{MAC Service Data Unit}
\newacronym{mss}{MSS}{Maximum Segment Size}
\newacronym{mtd}{MTD}{Machine-Type Device}
\newacronym{mtu}{MTU}{Maximum Transmission Unit}
\newacronym{nav}{NAV}{Network Allocation Vector}
\newacronym{ncbr}{NCBR}{Non-Constant Bitrate}
\newacronym{nfv}{NFV}{Network Function Virtualization}
\newacronym{nlos}{NLoS}{Non-Line-of-Sight}
\newacronym{nr}{NR}{New Radio}
\newacronym{nrmse}{NRMSE}{Normalized Root Mean Square Error}
\newacronym{ns3}{ns-3}{Network Simulator 3}
\newacronym{nsa}{NSA}{Non Stand Alone}
\newacronym{o2i}{O2I}{Outdoor-to-Indoor}
\newacronym{ofdm}{OFDM}{Orthogonal Frequency Division Multiplexing}
\newacronym{pa}{PA}{Position-aware}
\newacronym{pan}{PAN}{Personal Area Network}
\newacronym{pbch}{PBCH}{Physical Broadcast Channel}
\newacronym{pbss}{PBSS}{Personal Basic Service Set}
\newacronym{pcf}{PCF}{Point Coordinator Function}
\newacronym{pcp}{PCP}{\gls{pbss} Central Point}
\newacronym{pcpap}{PCP/AP}{\acrlong{pcp}/\acrlong{ap}}
\newacronym{pdcch}{PDCCH}{Physical Downlonk Control Channel}
\newacronym{pdcp}{PDCP}{Packet Data Convergence Protocol}
\newacronym{pdf}{PDF}{Probability Distribution Function}
\newacronym{pdsch}{PDSCH}{Physical Downlink Shared Channel}
\newacronym{per}{PER}{Packet Error Rate}
\newacronym{pdu}{PDU}{Packet Data Unit}
\newacronym{pf}{PF}{Proportional Fair}
\newacronym{pframe}{P-frame}{Predictive-coded frame}
\newacronym{pgw}{PGW}{Packet Gateway}
\newacronym{phy}{PHY}{Physical Layer}
\newacronym{ppdu}{PPDU}{PHY Protocol Data Unit}
\newacronym{ppp}{PPP}{Poisson Point Process}
\newacronym{prb}{PRB}{Physical Resource Block}
\newacronym{pss}{PSS}{Primary Synchronization Signal}
\newacronym{pucch}{PUCCH}{Physical Uplink Control Channel}
\newacronym{pusch}{PUSCH}{Physical Uplink Shared Channel}
\newacronym{qd}{QD}{Quasi Deterministic}
\newacronym{qoe}{QoE}{Quality of Experience}
\newacronym{qos}{QoS}{Quality of Service}
\newacronym{rach}{RACH}{Random Access Channel}
\newacronym{ran}{RAN}{Radio Access Network}
\newacronym[firstplural=Radio Access Technologies (RATs)]{rat}{RAT}{Radio Access Technology}
\newacronym{red}{RED}{Random Early Detection}
\newacronym{rf}{RF}{Radio Frequency}
\newacronym{rl}{RL}{Reinforcement Learning}
\newacronym{rlc}{RLC}{Radio Link Control}
\newacronym{rlf}{RLF}{Radio Link Failure}
\newacronym{rr}{RR}{Round Robin}
\newacronym{rrc}{RRC}{Radio Resource Control}
\newacronym{rrm}{RRM}{Radio Resource Management}
\newacronym{rs}{RS}{Remote Server}
\newacronym{rsrp}{RSRP}{Reference Signal Received Power}
\newacronym{rsrq}{RSRQ}{Reference Signal Received Quality}
\newacronym{rss}{RSS}{Received Signal Strength}
\newacronym{rssi}{RSSI}{Received Signal Strength Indicator}
\newacronym{rt}{RT}{Ray Tracer}
\newacronym{rts}{RTS}{Request to Send}
\newacronym{rtt}{RTT}{Round Trip Time}
\newacronym{rw}{RW}{Receive Window}
\newacronym{rx}{RX}{Receiver}
\newacronym{sa}{SA}{standalone}
\newacronym{sack}{SACK}{Selective Acknowledgment}
\newacronym{sap}{SAP}{Service Access Point}
\newacronym{sc}{SC}{Single Carrier}
\newacronym{sch}{SCH}{Secondary Cell Handover}
\newacronym{scm}{SCM}{Spatial Channel Model}
\newacronym{scoot}{SCOOT}{Split Cycle Offset Optimization Technique}
\newacronym{sdma}{SDMA}{Spatial Division Multiple Access}
\newacronym{sdr}{SDR}{Software Defined Radio}
\newacronym{semm}{SEMM}{SPCA-EDCA Mixed Mode}
\newacronym{si}{SI}{Study Item}
\newacronym{sib}{SIB}{Secondary Information Block}
\newacronym{sinr}{SINR}{Signal-to-Interference-plus-Noise Ratio}
\newacronym{sir}{SIR}{Signal-to-Interference Ratio}
\newacronym{sls}{SLS}{Sector-Level Sweep}
\newacronym{sm}{SM}{Saturation Mode}
\newacronym{snr}{SNR}{Signal-to-Noise Ratio}
\newacronym{son}{SON}{Self-Organizing Network}
\newacronym{sp}{SP}{Service Period}
\newacronym{spr}{SPR}{Service Period Request}
\newacronym{srs}{SRS}{Sounding Reference Signal}
\newacronym{ss}{SS}{Synchronization Signal}
\newacronym{sss}{SSS}{Secondary Synchronization Signal}
\newacronym{ssw}{SSW}{Sector Sweep}
\newacronym{sta}{STA}{Station}
\newacronym{stb}{STB}{Set Top Box}
\newacronym{tb}{TB}{Transport Block}
\newacronym[firstplural=Traffic Categories (TCs)]{tc}{TC}{Traffic Category}
\newacronym{tbtt}{TBTT}{Target Beacon Transmission Time}
\newacronym{tcp}{TCP}{Transmission Control Protocol}
\newacronym{tdd}{TDD}{Time Division Duplexing}
\newacronym{tdma}{TDMA}{Time Division Multiple Access}
\newacronym{tfl}{TfL}{Transport for London}
\newacronym{tgad}{TGad}{Task Group ad}
\newacronym{tgay}{TGay}{Task Group ay}
\newacronym{tsconst}{TSCONST}{Traffic Scheduling Constraint}
\newacronym{tsf}{TSF}{Timing Synchronization Function}
\newacronym{tm}{TM}{Transparent Mode}
\newacronym{trp}{TRP}{Transmitter Receiver Pair}
\newacronym{ts}{TS}{Traffic Stream}
\newacronym{tspec}{TSPEC}{Traffic Specification}
\newacronym{tti}{TTI}{Transmission Time Interval}
\newacronym{ttt}{TTT}{Time-to-Trigger}
\newacronym{tx}{TX}{Transmitter}
\newacronym[firstplural=Transmission Opportunities (TXOPs)]{txop}{TXOP}{Transmission Opportunity}
\newacronym{udp}{UDP}{User Datagram Protocol}
\newacronym{ue}{UE}{User Equipment}
\newacronym{ul}{UL}{Uplink}
\newacronym{um}{UM}{Unacknowledged Mode}
\newacronym{uma}{UMa}{Urban Macro}
\newacronym{uml}{UML}{Unified Modeling Language}
\newacronym{up}{UP}{User Priority}
\newacronym{utc}{UTC}{Urban Traffic Control}
\newacronym{vbr}{VBR}{Variable Bit Rate}
\newacronym{vm}{VM}{Virtual Machine}
\newacronym{vr}{VR}{Virtual Reality}
\newacronym{wbf}{WBF}{Wired Bias Function}
\newacronym{wf}{WF}{Wired-first}
\newacronym{wifi}{Wi-Fi}{Wireless Fidelity}
\newacronym{wigig}{WiGig}{Wireless Gigabit}
\newacronym{wlan}{WLAN}{Wireless Local Area Network}
\newacronym{xr}{XR}{eXtended Reality}

\makeglossaries
\glsdisablehyper

\newcommand{\ml}[1]{}
\newcommand{\md}[1]{}
\newcommand{\AZ}[1]{}

\newcommand{\TBI}{\ensuremath{T_{\rm BI}}}

\newcommand{\p}[1]{\ensuremath{p_{#1}}}
\newcommand{\Tp}[1]{\ensuremath{T_p^{#1}}}
\newcommand{\Tmin}[1]{\ensuremath{T_{\rm min}^{#1}}}
\newcommand{\Tmax}[1]{\ensuremath{T_{\rm max}^{#1}}}

\newcommand{\tstart}[1]{\ensuremath{t_{0, \rm start}^{#1}}}
\newcommand{\tend}[1]{\ensuremath{t_{\rm end}^{#1}}}
\newcommand{\Tblk}[1]{\ensuremath{T_{\rm blk}^{#1}}}

\newcommand{\tfeasmn}[2]{\ensuremath{t_{#1, \rm feas}^{#2}}}
\newcommand{\tlimmn}[2]{\ensuremath{t_{#1, \rm lim}^{#2}}}
\newcommand{\tfeas}{\ensuremath{t_{\rm feas}}}
\newcommand{\tlim}{\ensuremath{t_{\rm lim}}}
\newcommand{\tmin}{\ensuremath{t_{\rm min}}}
\newcommand{\tmax}{\ensuremath{t_{\rm max}}}

\theoremstyle{plain}

\usepackage[capitalize]{cleveref}
\crefname{section}{Sec.}{Secs.}
\crefname{property}{Property}{Properties}
\crefname{corollary}{Corollary}{Corollaries}
\crefname{feasibility}{Feasibility}{Feasibilities}

\glsunset{wifi}

\IEEEoverridecommandlockouts
\begin{document}

\addtolength{\topmargin}{+0.06cm}

\title{Exploiting Scheduled Access Features\\of mmWave WLANs for Periodic Traffic Sources}

\author{
 \IEEEauthorblockN{Mattia Lecci, Matteo Drago, Andrea Zanella, Michele Zorzi}
 \IEEEauthorblockA{\textit{Department of Information Engineering}, \textit{University of Padova}, Italy\\
  E-mails: \texttt{\{name.surname\}@dei.unipd.it}}
 \thanks{This work was partially supported by NIST under Award No. 60NANB19D122.
  Mattia Lecci's activities were supported by \textit{Fondazione CaRiPaRo} under the grant ``Dottorati di Ricerca 2018.''}
}

\maketitle

\begin{abstract}
 Many current and future multimedia and industrial applications, like video streaming, eXtended Reality or remote robot control, are characterized by periodic data transmissions with strict latency and reliability constraints.
 In an effort to meet the stringent demand of such traffic sources, the \acrshort{wigig} standards support a contention-free channel access mechanism, named Service Period, that makes it possible to allocate dedicated time intervals to certain wireless stations.
 However, the standard only covers the fundamental aspects that ensure interoperability, while the actual schedule logic is left to vendors.

 In this paper, we propose two algorithms for joint admission control and scheduling of periodic traffic streams with contrasting performance objectives, specifically a \textit{simple scheduler} and a \textit{max-min fair scheduler}.
 The schemes are compared in two different scenarios, in order to characterize and highlight some fundamental trade-offs.
 As expected from their design principles, the simple scheduler tends to trade acceptance rate for resource availability, contrary to the max-min fair scheduler, giving to implementers a clear performance trade-off, although performance cannot be balanced by means of a tunable parameter.

\end{abstract}


\IEEEpeerreviewmaketitle

\begin{tikzpicture}[remember picture,overlay]
\node[anchor=north,yshift=-10pt] at (current page.north) {\parbox{\dimexpr\textwidth-\fboxsep-\fboxrule\relax}{
\centering\footnotesize This paper has been submitted to IEEE MedComNet 2021. Copyright may change without notice.}};
\end{tikzpicture}

\glsresetall
\glsunset{wifi}

\section{Introduction} 
\label{sec:introduction}

The always-increasing capacity of wireless systems is promoting the design of applications and services with increasingly challenging demands, such as video streaming, teleconference, telepresence, \acrfull{xr}, among others~\cite{tgayUsageModel}.
In order to meet the demand in terms of data rate of such applications, the latest versions of the \gls{wifi} standard, i.e., IEEE~802.11ad and 802.11ay, also known as \gls{wigig}, offer the possibility to communicate over the mmWave band at 60~GHz, where multiple 2.16~GHz channels are available.
By taking advantage of techniques such as channel bonding and \gls{mimo}, and by introducing novel features to the protocol stack, these standards can provide data rates over 100~Gbps~\cite{11ayMagazine}.

However, many applications also have very stringent \gls{qos} requirements, in particular in terms of delay and jitter, which may be incompatible with the stochastic nature of contention-based channel access mechanisms generally supported by legacy \glspl{wlan}.
To address this problem, the \gls{wigig} standards introduced a contention-free access mechanism that allows a \gls{sta} to reserve radio resources at regular time intervals.
These resources are organized in blocks, called \glspl{sp}, and the standards only specify the basic procedures to ensure inter-vendor compatibility, leaving the design and the implementation of the scheduler to the manufacturers.

Regarding the practical design, handling multiple periodic traffic streams can be problematic, especially when traffic flows with different periodicities coexist.
In this case, it is necessary to anticipate collisions among different periodic allocations and either adjust them or, in the worst case, reject new incompatible requests.
Furthermore, even if a collection of requests with identical traffic requirements is considered, upon receiving a new request, the scheduler needs to decide whether to rearrange the previously allocated resources to improve fairness and efficiency, or to maintain the original schedule and then best accommodate the new request, in order not to perturb the pre-existing streams but potentially reaching suboptimal resource allocation.
Moreover, \glspl{sp} are subject to a number of constraints, described in \cref{sec:framework_description}, which need to be accounted for when designing and optimizing scheduling algorithms.

With these challenges in mind, in this work we address both admission control and resource allocation for multiple periodic traffic sources, following the constraints given by the \gls{wigig} standards.
Specifically, we cast the periodic scheduling problem within the \gls{wigig} allocation framework and design a simple and efficient algorithm to check for the feasibility of a new request. 
We then propose a simple admission control algorithm with limited scheduling capabilities, as well as a more elaborate and optimized strategy to increase the admission rate and, possibly, the fairness among independent flows.
Finally, we compare these two policies and shed some light on basic trade-offs.

The rest of the paper is organized as follows.
The resource allocation framework is described in \cref{sec:framework_description}, while \cref{sec:state_of_the_art} provides an overview of the State of the Art on related problems, and motivates our need to fill the gap of the current literature in this topic.
Then, we present the proposed algorithms in \cref{sec:scheduling_algorithms}. Performance analysis is presented in \cref{sec:results}.
Finally, in \cref{sec:conclusions} we draw our conclusions and propose possible extensions of this work.


\section{Framework Description} 
\label{sec:framework_description}
Based on~\cite{standard802.11_2016}, \glspl{sta} can request the \gls{ap} to reserve periodic transmission intervals by sending a control frame containing the required periodicity ($\p{}$) and the minimum and maximum duration of each allocation ($[\Tmin{}, \Tmax{}]$).
The \gls{ap} advertises the allocated \glspl{sp} to the \glspl{sta} at each \gls{bi}, specifying the starting time, duration, and periodicity of each block.
The allocation needs to comply with a number of constraints:
\begin{enumerate}
 \item Periodicity ($\p{}$) can only be an integer multiple ($\p{} \in \mathbb{N}$) or an integer fraction ($\p{}^{-1} \in \mathbb{N}$) of a \gls{bi} (\TBI), thus the block periodicity interval will be $\Tp{} = \p{} \, \TBI $;
 \item Allocation blocks cannot be scheduled across the \gls{bi} boundaries;
 \item The allocated block duration $\Tblk{}$ should fall in the range $[\Tmin{}, \Tmax{}]$ specified in the resource request.
\end{enumerate}
A more detailed description of the constraints imposed by the standard can be found in~\cite{standard802.11_2016} and~\cite{mohebi2020challenges}.

Since this work is focused on  allocation algorithms for periodic traffic sources, we neglect the \glspl{cbap}, which is present in each \gls{bi} for asynchronous traffic.
In addition, to compare the scheduling algorithms in challenging conditions, we assume that the allocated resources will be maintained indefinitely, so that the channel load increases progressively as new resource reservations are accepted.
For the sake of simplicity and clarity, we also assume that the allocation blocks of a given accepted request are not fractioned into multiple disjoint intervals (i.e., each \gls{sp} will consist of a single time interval of duration $\Tblk{}$).
Furthermore, we consider a \textit{strict periodicity} constraint, which prevents the scheduler from changing the starting time of already allocated blocks, while the block duration $\Tblk{}$ can be freely changed within the interval $\qty[\Tmin{}, \Tmax{}]$.


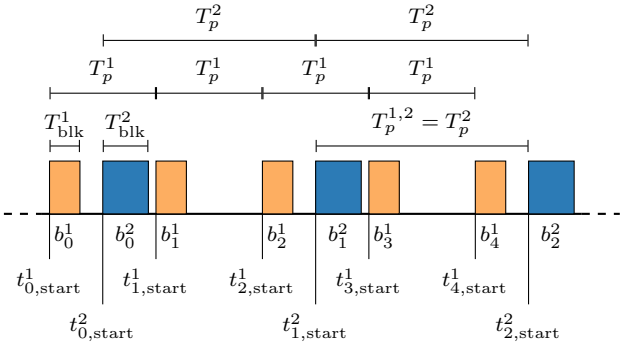
\begin{figure}[tp]
 \centering
 \newcommand\fscale{1}
 \begin{tikzpicture}[scale=\fscale, every node/.style={transform shape, font=\footnotesize}]

\definecolor{color0}{rgb}{0.1725,0.4824,0.7137}
\definecolor{color1}{rgb}{0.9922,0.6824,0.3804}

\def\TstartOne{0.1}
\def\tblkone{0.4}
\def\TPone{1.4}
\def\NmaxAllocOne{4}

\def\TstartTwo{0.8}
\def\tblktwo{0.6}
\def\TPtwo{2.8}
\def\NmaxAllocTwo{2}

\def\allocHeight{.7}

\pgfmathsetmacro{\NmaxAllocOneMinusOne}{\NmaxAllocOne-1}
\pgfmathsetmacro{\NmaxAllocTwoMinusOne}{\NmaxAllocTwo-1}

\draw[thick, dashed] (-.5,0) -- (0,0);
\draw[thick] (0,0) -- (7.1,0);
\draw[thick, dashed] (7.1,0) -- (7.6,0);

\foreach \x in {0,1,...,\NmaxAllocOne}
  {
    \coordinate (A\x) at ($(\TstartOne,0) + \x*(\TPone,0)$) {};
    \draw[fill=color1] (A\x) rectangle ++(\tblkone,\allocHeight) node {};
    \node[below] at ($(A\x) + (\tblkone / 2, 0)$) {$b_\x^1$};
  }

\foreach \x in {0,1,...,\NmaxAllocTwo}
  {
    \coordinate (B\x) at ($(\TstartTwo,0) + \x*(\TPtwo,0)$) {};
    \draw[fill=color0] (B\x) rectangle ++(\tblktwo,\allocHeight) node {};
    \node[below] at ($(B\x) + (\tblktwo / 2, 0)$) {$b_\x^2$};
  }

\draw[|-|] (\TstartOne, \allocHeight+0.2) -- node[above] {$\Tblk{1}$} ++(\tblkone, 0);
\draw[|-|] (\TstartTwo, \allocHeight+0.2) -- node[above] {$\Tblk{2}$} ++(\tblktwo, 0);
\draw[|-|] (\TstartTwo + \TPtwo, \allocHeight+0.2) -- node[above] {$\Tp{1,2} = \Tp{2}$} ++(\TPtwo, 0);

\foreach \x in {0,1,...,\NmaxAllocOneMinusOne}
  {
    \draw[|-|] (\TstartOne + \x * \TPone, \allocHeight+0.9) -- node[above] {$\Tp{1}$} ++(\TPone, 0);
  }

\foreach \x in {0,1,...,\NmaxAllocTwoMinusOne}
  {
    \draw[|-|] (\TstartTwo + \x * \TPtwo, \allocHeight+1.6) -- node[above] {$\Tp{2}$} ++(\TPtwo, 0);
  }

\draw (A0) -- ++(0,-0.6) node[below] {$t_{0, \rm start}^1$};
\foreach \x in {1,...,\NmaxAllocOne}
  {
    \draw (A\x) -- ++(0,-0.6) node[below] {$t_{\x, \rm start}^1$};
  }

\draw (B0) -- ++(0,-1.2) node[below] {$t_{0, \rm start}^2$};
\foreach \x in {1,...,\NmaxAllocTwo}
  {
    \draw (B\x) -- ++(0,-1.2) node[below] {$t_{\x, \rm start}^2$};
  }

\end{tikzpicture}
 \caption{Example of allocations $A_1$ (orange) and $A_2$ (blue), with $\p{2}=2\p{1}$.}
 \label{fig:joint_alloc}
\end{figure}

\section{State of the Art} 
\label{sec:state_of_the_art}

Many works analyzing the \gls{mac} layer of the \gls{wigig} standards focus mostly on \glspl{cbap}, which extend the traditional WiFi access to cope with directional communications, either neglecting \gls{sp} allocations or considering extremely simple allocation schemes.
In \cite{babich2009throughput}, the authors proposed a mathematical framework for the analysis of \gls{e2e} metrics in 802.11-based systems, comparing throughput and average packet delay in scenarios where the nodes are equipped with advanced antenna systems.
The characteristics of the \gls{dcf} were taken into account, for which a theoretical performance analysis was carried out in \cite{bianchi2000performance}.
Instead, the authors in~\cite{pielli2019analytical} present a model to assess the performance of \glspl{cbap} for the IEEE 802.11ad standard, taking into account a directional channel model and the presence of scheduled \glspl{sp}, but they do not focus on how to assign such \glspl{sp}.

To the best of our knowledge, little work has been done on contention-free scheduling for \gls{wigig} networks.
In~\cite{hemanth2013performance,rajan2016saturation}, the authors analyze the case where all contention-free allocations occupy the beginning of each \gls{bi}, while the rest of the interval is left for a single \gls{cbap}.
This allocation strategy, however, cannot support requests for periodic resource allocations with time period shorter than $\TBI{}$.
The authors of~\cite{khorov2016mathematical}, instead, propose an accurate mathematical analysis of the performance of a realistic \gls{vbr} traffic source in the presence of channel errors, when using a periodic resource allocation scheme, but do not tackle the problem of scheduling multiple periodic allocations at once.

On the other hand, the problem of periodic scheduling has been widely studied in other areas, such as real-time computation and task scheduling, where the objective is to complete tasks within a given time, while minimizing the resource utilization.
For example, the authors of~\cite{liu73scheduling} develop and compare heuristic algorithms for scheduling tasks with hard periodic deadlines and constant resource utilization, showing that a deadline-first approach ensures maximum resource utilization.
In~\cite{ramamritham95allocation}, the authors try to schedule safety-critical periodic tasks with precedence constraints, distributed over multi-processor systems using an adapted deadline-first approach, while the authors of~\cite{cheng95allocation} use simulated annealing to optimize a similar problem.
Finally, \cite{zhu03multiple} finds a low-overhead optimal solution (from a resource utilization point of view) assuming that tasks have a fixed resource requirement.

All these approaches, however, cannot be directly used in \gls{wigig} systems, either because they are not compliant with the constraints imposed by the resource allocation procedures (i.e., granularity of the allocation periods, \gls{bi} boundaries), or because they cannot exploit the \gls{wigig} standards' flexibility (e.g., the dynamic allocation of $\Tblk{}$). This work contributes to fill the gap by proposing admission control and scheduling algorithms that account for the specific features of \gls{mmw} \glspl{wlan}.


\section{Scheduling Algorithms} 
\label{sec:scheduling_algorithms}
\begin{figure}[t!]
 \newcommand\fscale{1}
 \centering
 \begin{subfigure}[b]{\columnwidth}
  \centering
  \begin{tikzpicture}[scale=\fscale, every node/.style={transform shape, font=\footnotesize}]

\definecolor{color0}{rgb}{0.1725,0.4824,0.7137}
\definecolor{color1}{rgb}{0.9922,0.6824,0.3804}

\def\tbi{3}

\def\TstartOne{0}
\def\tblkone{0.4}
\def\TPone{1.5}
\def\NminAllocOne{-1}
\def\NmaxAllocOne{2}

\def\TstartTwo{0}
\def\tblktwo{0.25}
\def\TPtwo{1}
\def\NminAllocTwo{0}
\def\NmaxAllocTwo{3}

\def\allocHeight{.7}

\draw[thick, dashed] (-2.5,0) -- (-2,0);
\draw[thick] (-2,0) -- (4,0);
\draw[thick, dashed] (4,0) -- (4.5,0);

\foreach \x in {\NminAllocOne,...,\NmaxAllocOne}
  {
    \coordinate (A\x) at ($(\TstartOne,0) + \x*(\TPone,0)$) {};
    \draw[fill=color0] (A\x) rectangle ++(\tblkone,\allocHeight) node {};
  }

\foreach \x in {\NminAllocTwo,...,\NmaxAllocTwo}
  {
    \coordinate (A\x) at ($(\TstartTwo,0) + \x*(\TPtwo,0)$) {};
    \draw[fill=color1] (A\x) rectangle ++(\tblktwo,\allocHeight) node {};
  }

\draw[pattern color=color0, pattern=north west lines] (0,0) rectangle ++(\tblkone,\allocHeight) node {};
\draw[pattern color=color0, pattern=north west lines] (\tbi,0) rectangle ++(\tblkone,\allocHeight) node {};

\draw[fill=green] (0,0) rectangle (\TPtwo,-0.1) node {};
\draw (\TPtwo,0) -- ++(0,-.2) node[below right] {$\tmax$};

\draw[|-|] (0,\allocHeight+0.1) -- node[above] {$\Delta t$} ++(\tblkone, 0);
\draw[|-|] (0,\allocHeight+0.6) -- node[above] {$\Tp{1,2}$} ++(\tbi, 0);

\draw (0,0) -- ++(0,-.2) node[below] {$\tstart{2} = \tmin$};

\end{tikzpicture}
  \caption{Scheduling step 0.}
  \label{fig:feasibility_check_fail_0}
 \end{subfigure}
 \\
 \begin{subfigure}[b]{\columnwidth}
  \centering
  \begin{tikzpicture}[scale=\fscale, every node/.style={transform shape, font=\footnotesize}]

\definecolor{color0}{rgb}{0.1725,0.4824,0.7137}
\definecolor{color1}{rgb}{0.9922,0.6824,0.3804}

\def\tbi{3}

\def\TstartOne{0}
\def\tblkone{0.4}
\def\TPone{1.5}
\def\NminAllocOne{-1}
\def\NmaxAllocOne{2}

\def\TstartTwo{\tblkone}
\def\tblktwo{0.25}
\def\TPtwo{1}
\def\NminAllocTwo{0}
\def\NmaxAllocTwo{3}

\def\allocHeight{.7}

\draw[thick, dashed] (-2.5,0) -- (-2,0);
\draw[thick] (-2,0) -- (4,0);
\draw[thick, dashed] (4,0) -- (4.5,0);

\foreach \x in {\NminAllocOne,...,\NmaxAllocOne}
  {
    \coordinate (A\x) at ($(\TstartOne,0) + \x*(\TPone,0)$) {};
    \draw[fill=color0] (A\x) rectangle ++(\tblkone,\allocHeight) node {};
  }

\foreach \x in {\NminAllocTwo,...,\NmaxAllocTwo}
  {
    \coordinate (A\x) at ($(\TstartTwo,0) + \x*(\TPtwo,0)$) {};
    \draw[fill=color1] (A\x) rectangle ++(\tblktwo,\allocHeight) node {};
  }

\draw[pattern color=color0, pattern=north west lines] (\TstartOne+\TPone,0) rectangle (\TstartTwo+\TPtwo+\tblktwo,\allocHeight);

\draw[fill=red] (0,0) rectangle (\tblkone,-0.1);
\draw[fill=green] (\tblkone,0) rectangle (\TPtwo,-0.1);
\draw (\TPtwo,0) -- ++(0,-.2) node[below] {$\tmax$};

\draw[|->] (0,\allocHeight+0.1) -- node[above] {$\tstart{2}$} ++(\tblkone,0);
\draw[|-|] (\TstartTwo+\TPtwo,\allocHeight+0.1) -- node[above] {$\Delta t$} (\TstartOne+\TPone+\tblkone, \allocHeight+0.1);

\draw (0,0) -- ++(0,-.2) node[below] {$\tmin$};

\end{tikzpicture}
  \caption{Scheduling step 1.}
  \label{fig:feasibility_check_fail_1}
 \end{subfigure}
 \\
 \begin{subfigure}[b]{\columnwidth}
  \centering
  \begin{tikzpicture}[scale=\fscale, every node/.style={transform shape, font=\footnotesize}]

\definecolor{color0}{rgb}{0.1725,0.4824,0.7137}
\definecolor{color1}{rgb}{0.9922,0.6824,0.3804}

\def\tbi{3}

\def\TstartOne{0}
\def\tblkone{0.4}
\def\TPone{1.5}
\def\NminAllocOne{-1}
\def\NmaxAllocOne{2}

\def\tblktwo{0.25}
\def\TPtwo{1}
\def\TstartTwo{\TstartOne+\TPone+\tblkone-\TPtwo}
\def\NminAllocTwo{0}
\def\NmaxAllocTwo{2}

\def\allocHeight{.7}

\draw[thick, dashed] (-2.5,0) -- (-2,0);
\draw[thick] (-2,0) -- (4,0);
\draw[thick, dashed] (4,0) -- (4.5,0);

\foreach \x in {\NminAllocOne,...,\NmaxAllocOne}
  {
    \coordinate (A\x) at ($(\TstartOne,0) + \x*(\TPone,0)$) {};
    \draw[fill=color0] (A\x) rectangle ++(\tblkone,\allocHeight) node {};
  }

\foreach \x in {\NminAllocTwo,...,\NmaxAllocTwo}
  {
    \coordinate (A\x) at ($(\TstartTwo,0) + \x*(\TPtwo,0)$) {};
    \draw[fill=color1] (A\x) rectangle ++(\tblktwo,\allocHeight) node {};
  }

\draw[pattern color=color0, pattern=north west lines] (0,0) rectangle ++(\tblkone,\allocHeight) node {};
\draw[pattern color=color0, pattern=north west lines] (\tbi,0) rectangle ++(\tblkone,\allocHeight) node {};

\draw[fill=red] (0,0) rectangle (\TstartTwo,-0.1) node {};
\draw[fill=green] (\TstartTwo,0) rectangle (\TPtwo,-0.1) node {};

\draw (\TPtwo,0) -- ++(0,-.2);
\draw (\TstartTwo+\tblktwo,0) -- ++(0,-.2) node[below] {$t_{0,\rm end}^{2} > \tmax$};

\draw[|->] (\tblkone,\allocHeight+0.1) -- node[above] {$\tstart{2}$} (\TstartTwo,\allocHeight+0.1);

\draw (0,0) -- ++(0,-.2) node[below left] {$\tmin$};

\end{tikzpicture}
  \caption{Scheduling step 2.}
  \label{fig:feasibility_check_fail_2}
 \end{subfigure}
 \caption{Feasibility check for an \textit{infeasible} pair of allocations, where $A_1$ (blue) was a pre-existing allocation with $\p{1}=\frac{1}{2}$, and the algorithm is checking whether a new allocation $A_2$ (orange) with $\p{2}=\frac{1}{3}$ is compatible.}
 \label{fig:feasibility_check_fail}
\end{figure}
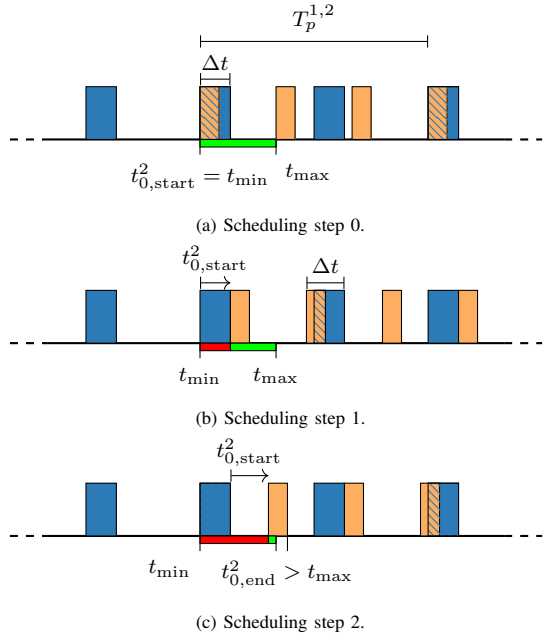

\begin{algorithm}[t!]
 \footnotesize
 \caption{Feasibility check under strong periodicity conditions (see \cref{fig:feasibility_check_fail,fig:feasibility_check_succ}).}
 \label{alg:strict_periodicity_feasibility}
 \begin{algorithmic}[1]
  \REQUIRE $\qty{A_1,\ldots, A_{N-1}}$ (fixed), $A_N$ (new allocation), $\qty[\tmin, \tmax]$

  \STATE Compute $\Tp{1,\ldots,N}$
  \STATE $\tstart{N} = t_{0, \rm start}^N \leftarrow \tmin$ \COMMENT{\cref{fig:feasibility_check_fail_0}}

  \WHILE{$t_{0, \rm end}^{N} < \tmax$}
  \STATE Check for collisions in $\left[\tmin, \tmin + \Tp{1,\ldots,N} \right)$
  \IF{no collisions}
  \STATE $\tfeas \leftarrow \tstart{N}$
  \RETURN $A_N$ is feasible with starting time $\tfeas$ \COMMENT{\cref{fig:feasibility_check_succ}}
  \ELSE
  \STATE Allocation block $h \in A_N$ collides with allocation block $k \in A_i$, for some $i \in {1,\ldots,N-1}$
  \STATE $\Delta t \leftarrow t_{k, \rm end}^i - t_{h, \rm start}^N$
  \STATE $\tstart{N} \leftarrow \tstart{N} + \Delta t$
  \ENDIF
  \ENDWHILE

  \RETURN $A_N$ is not a feasible allocation \COMMENT{\cref{fig:feasibility_check_fail}}
 \end{algorithmic}
\end{algorithm}

We denote by $A_n= (\tstart{n}, \Tp{n}, \Tblk{n})$ the allocation for the $n$-th traffic stream, where $\tstart{n}$ is the starting epoch, $\Tp{n}$ is the period, and $ \Tblk{n}$ is the allocated duration of each individual block.
Therefore, the allocation consists of a sequence of blocks, where the $k$-th block of the $n$-th traffic stream takes the interval $b_k^n = \qty(t_{k, \rm start}^n, t_{k, \rm end}^{n})$, where
\begin{equation}
 \begin{aligned}\label{eq:alloc_blk_start_end}
  t_{k, \rm start}^n & = \tstart{n} + k \Tp{n}           \;;  \\
  t_{k, \rm end}^{n} & = \tstart{n} + k \Tp{n} + \Tblk{n} \;;
 \end{aligned}
\end{equation}
for $k=0,1,2,\ldots\;$.
A graphical example is shown in \cref{fig:joint_alloc}.

Following this definition we can say that, given $N$ distinct allocations $A_1,\ldots,A_N$, they are jointly periodic over a period
\begin{equation}
  T_p^{1,\ldots,N} = \mathrm{lcm}\left(T_p^i, \ldots, T_p^N \right),
\end{equation}
where $\mathrm{lcm}$ indicates the \textit{least common multiple} of the periods.
Note that, since all block periods are integer multiples or fractions of \TBI, the \gls{lcm} can always be properly defined~\cite{zazkis2015lcm} as
\begin{equation}
  \mathrm{lcm}\left( \frac{a}{b}, \frac{c}{d} \right) = \frac{\mathrm{lcm}(a,c)}{\mathrm{gcf}(b,d)},
\end{equation}
where $\mathrm{gfc}$ is the \textit{greatest common factor}.

Given the periodicity of the allocation patterns, a new allocation $A_N$ should start within a time interval $\Tp{N}$ since the beginning of the \gls{bi}.
Moreover, a necessary requirement for admission is that in an interval of duration $T_P^{1,\ldots,N}$, no block $b_h^N$ overlaps with any block $b_k^n$, $n \in \{1,,\ldots,N-1\}$, $\forall h,k\geq 0$.

The remainder of this section is structured as follows: in \cref{sub:feasibility_check_algorithm} we will illustrate an algorithm to efficiently check whether a new allocation is compatible with a pre-existing schedule, in \cref{sub:simple_scheduler} we will present a simple scheduling algorithm, and finally in \cref{sub:max_min_fair_scheduler} we will describe in detail a more complex algorithm that aims at minimizing the rejection of new allocations under the \textit{strict periodicity} assumption.

\subsection{Feasibility Check Algorithm} 
\label{sub:feasibility_check_algorithm}

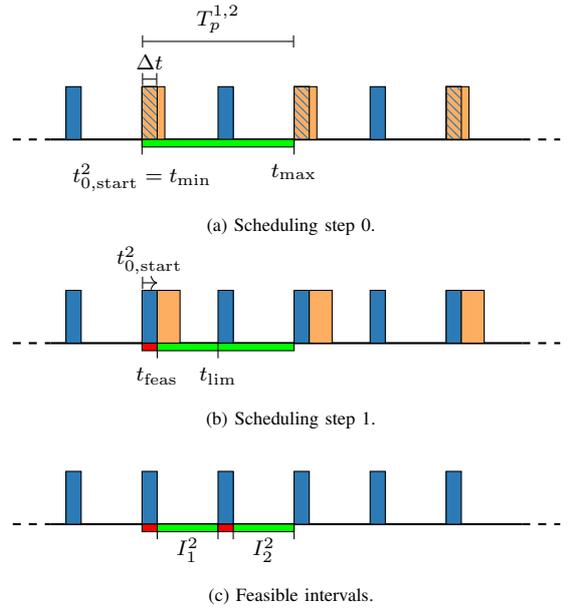
\begin{figure}[t!]
 \newcommand\fscale{1}
 \centering
 \begin{subfigure}[b]{\columnwidth}
  \centering
  \begin{tikzpicture}[scale=\fscale, every node/.style={transform shape, font=\footnotesize}]

\definecolor{color0}{rgb}{0.1725,0.4824,0.7137}
\definecolor{color1}{rgb}{0.9922,0.6824,0.3804}

\def\TstartOne{0}
\def\tblkone{0.2}
\def\TPone{1}
\def\NminAllocOne{-1}
\def\NmaxAllocOne{4}

\def\TstartTwo{0}
\def\tblktwo{0.3}
\def\TPtwo{2}
\def\NminAllocTwo{0}
\def\NmaxAllocTwo{2}

\def\allocHeight{.7}

\draw[thick, dashed] (-1.7,0) -- (-1.2,0);
\draw[thick] (-1.2,0) -- (5,0);
\draw[thick, dashed] (5,0) -- (5.5,0);

\foreach \x in {\NminAllocOne,...,\NmaxAllocOne}
  {
    \coordinate (A\x) at ($(\TstartOne,0) + \x*(\TPone,0)$) {};
    \draw[fill=color0] (A\x) rectangle ++(\tblkone,\allocHeight) node {};
  }

\foreach \x in {\NminAllocTwo,...,\NmaxAllocTwo}
  {
    \coordinate (A\x) at ($(\TstartTwo,0) + \x*(\TPtwo,0)$) {};
    \draw[fill=color1] (A\x) rectangle ++(\tblktwo,\allocHeight) node {};
  }

\foreach \x in {\NminAllocTwo,...,\NmaxAllocTwo}
  {
    \coordinate (A\x) at ($(\TstartTwo,0) + \x*(\TPtwo,0)$) {};
    \draw[pattern color=color0, pattern=north west lines] (A\x) rectangle ++(\tblkone,\allocHeight) node {};
  }

\draw[fill=green] (0,0) rectangle (\TPtwo,-0.1) node {};
\draw (\TPtwo,0) -- ++(0,-.2) node[below] {$\tmax$};

\draw[|-|] (0,\allocHeight+0.1) -- node[above] {$\Delta t$} ++(\tblkone, 0);
\draw[|-|] (0,\allocHeight+0.6) -- node[above] {$\Tp{1,2}$} ++(\TPtwo, 0);

\draw (0,0) -- ++(0,-.2) node[below] {$\tstart{2} = \tmin$};

\end{tikzpicture}
  \caption{Scheduling step 0.}
  \label{fig:feasibility_check_succ_0}
 \end{subfigure}
 \\
 \begin{subfigure}[b]{\columnwidth}
  \centering
  \begin{tikzpicture}[scale=\fscale, every node/.style={transform shape, font=\footnotesize}]

\definecolor{color0}{rgb}{0.1725,0.4824,0.7137}
\definecolor{color1}{rgb}{0.9922,0.6824,0.3804}

\def\TstartOne{0}
\def\tblkone{0.2}
\def\TPone{1}
\def\NminAllocOne{-1}
\def\NmaxAllocOne{4}

\def\TstartTwo{\tblkone}
\def\tblktwo{0.3}
\def\TPtwo{2}
\def\NminAllocTwo{0}
\def\NmaxAllocTwo{2}

\def\allocHeight{.7}

\draw[thick, dashed] (-1.7,0) -- (-1.2,0);
\draw[thick] (-1.2,0) -- (5,0);
\draw[thick, dashed] (5,0) -- (5.5,0);

\foreach \x in {\NminAllocOne,...,\NmaxAllocOne}
  {
    \coordinate (A\x) at ($(\TstartOne,0) + \x*(\TPone,0)$) {};
    \draw[fill=color0] (A\x) rectangle ++(\tblkone,\allocHeight) node {};
  }

\foreach \x in {\NminAllocTwo,...,\NmaxAllocTwo}
  {
    \coordinate (A\x) at ($(\TstartTwo,0) + \x*(\TPtwo,0)$) {};
    \draw[fill=color1] (A\x) rectangle ++(\tblktwo,\allocHeight) node {};
  }

\draw[fill=red] (0,0) rectangle (\tblkone,-0.1);
\draw[fill=green] (\tblkone,0) rectangle (\TPtwo,-0.1);

\draw (\tblkone,0) -- ++(0,-.2) node[below] {$\tfeas$};
\draw (\TPone,0) -- ++(0,-.2) node[below] {$\tlim$};

\draw[|->] (0,\allocHeight+0.1) -- node[above] {$\tstart{2}$} ++(\tblkone,0);

\end{tikzpicture}
  \caption{Scheduling step 1.}
  \label{fig:feasibility_check_succ_1}
 \end{subfigure}
 \vspace{0.1pt}
 \\
 \begin{subfigure}[b]{\columnwidth}
  \centering
  \begin{tikzpicture}[scale=\fscale, every node/.style={transform shape, font=\footnotesize}]

\definecolor{color0}{rgb}{0.1725,0.4824,0.7137}
\definecolor{color1}{rgb}{0.9922,0.6824,0.3804}

\def\TstartOne{0}
\def\tblkone{0.2}
\def\TPone{1}
\def\NminAllocOne{-1}
\def\NmaxAllocOne{4}

\def\TstartTwo{\tblkone}
\def\tblktwo{0.3}
\def\TPtwo{2}
\def\NminAllocTwo{0}
\def\NmaxAllocTwo{2}

\def\allocHeight{.7}

\draw[thick, dashed] (-1.7,0) -- (-1.2,0);
\draw[thick] (-1.2,0) -- (5,0);
\draw[thick, dashed] (5,0) -- (5.5,0);

\foreach \x in {\NminAllocOne,...,\NmaxAllocOne}
  {
    \coordinate (A\x) at ($(\TstartOne,0) + \x*(\TPone,0)$) {};
    \draw[fill=color0] (A\x) rectangle ++(\tblkone,\allocHeight) node {};
  }

\draw[fill=green] (\tblkone,0) rectangle node[below] {$I_1^2$} (\TPone,-0.1);
\draw (\TstartTwo,0) -- ++(0,-.2);
\draw (\TPone,0) -- ++(0,-.2);

\draw[fill=green] (\TPone+\tblkone,0) rectangle node[below] {$I_2^2$} (\TPtwo,-0.1);
\draw (\TPone+\tblkone,0) -- ++(0,-.2);
\draw (\TPtwo,0) -- ++(0,-.2);

\draw[fill=red] (0,0) rectangle (\tblkone,-0.1);
\draw[fill=red] (\TPone,0) rectangle (\TPone+\tblkone,-0.1);

\end{tikzpicture}
  \caption{Feasible intervals.}
  \label{fig:feasibility_check_succ_2}
 \end{subfigure}
 \caption{Feasibility check for a \textit{feasible} pair of allocations, where $A_1$ (blue) was a pre-existing allocation with $\p{1}=\frac{1}{4}$, and the algorithm is checking whether a new allocation $A_2$ (orange) with $\p{2}=\frac{1}{2}$ is compatible.}
 \label{fig:feasibility_check_succ}
\end{figure}

This feasibility check can be performed as described in \cref{alg:strict_periodicity_feasibility}, whose arguments consist of the existing allocations $A_1,\ldots,A_{N-1}$, the new request $A_N$ and a search interval $\qty[\tmin, \tmax]$.
For reasons that will be clear later, we assume that the existing allocations cannot be changed, while for $A_N$ only $\tstart{N}$ can be modified, keeping the period $T_p^N$ and the block duration $\Tblk{N}$ fixed.
Based on these input values, the aim of the procedure is to find the earliest feasible starting time $\tfeas^N$ such that a block of duration $\Tblk{N}$ fits in the search interval.

To do so, starting from $\tmin$, the algorithm progressively shifts the starting time by an interval $\Delta t$ (described in \cref{alg:strict_periodicity_feasibility}) until either all feasibility conditions are met, or $t_{0, \rm end}^{N} > \tmax$, in which case the allocation $A_N$ with block duration $\Tblk{N}$ is rejected.

A trivial example involves $A_1$, i.e., the first received allocation request from a \gls{sta}.
In this case, $\tmin$ will be set to the start time of the first \gls{bi} following the reception of the request, while $\tmax = \Tp{N}$ to guarantee the periodicity.
Since no previous allocated \glspl{sp} exist, $A_1$ is immediately accepted with $\tfeas = \tmin$.
It is important to highlight that, however, by choosing specific combinations of input parameters, \cref{alg:strict_periodicity_feasibility} can be used also by more advanced scheduling schemes, as explained later.

Given any feasible starting time $\tfeas$, it is useful to compute the rightmost \textit{boundary} of the allocation, i.e., the largest interval $[\tfeas,\tlim]$ that would still make $b_0^N \in [\tfeas,\tlim]$ and, in turn, $A_N$ feasible, even for larger values $\tstart{N}$ and $\Tblk{N}$.
This boundary can be computed by finding the minimum distance between each $b_h \in A_N$ and each $b_k \in A_n$, $n \neq N$.
The final results will be the minimum measured distance.
A graphical illustration of how the algorithm behaves when the new request is infeasible is shown in \cref{fig:feasibility_check_fail}, while a new feasible request is shown in \cref{fig:feasibility_check_succ}.

Following this definition and the above numerical example, the first allocation request to be generated, i.e., $A_1$, will find itself in the optimal condition where $\tfeas=\tmin$ and $\tlim=\tmax$.

In general, multiple feasible intervals exist.
To find an exhaustive list, we can iterate \cref{alg:strict_periodicity_feasibility} with $\tstart{N}$ initialized to the start time of the \gls{bi}, and progressively updated at each iteration with the value of $\tlim$ found in the previous execution.
This procedure continues until the shift of $\tstart{N}$ leads to an infeasible allocation. 
We define the list of feasible intervals (which depend on $\Tblk{N}$) as $\mathcal{I}_N = \qty{I_1^N,\ldots,I_M^N}$, where $I_m^N = \qty[\tfeasmn{m}{N}, \tlimmn{m}{N}]$, $m=1,\ldots,M$ (see \cref{fig:feasibility_check_succ_2}).
Hence, each of these intervals delimits the finite number of intervals in which the new allocation $A_N$ can be fitted, considering all previous allocations.
A good scheduling algorithm should then assess which interval yields the best overall performance, possibly trying to optimize a target \gls{kpi}.


\subsection{Simple Scheduler} 
\label{sub:simple_scheduler}
The first scheduler that we propose assumes that the block duration and periodicity of already accepted traffic streams cannot be varied.
Then, a new request $A_N$ with a block duration of $\Tblk{N} \in [\Tmin{N}, \Tmax{N}]$ can be accepted only if there exists a feasible interval in $\mathcal{I}_N $ with a duration of at least $\Tmin{N}$.
Therefore, the maximum amount of available resources that can be allocated to $A_N$ is determined by the longest feasible interval, or by $\Tmax{N}$, whichever is smaller; $\tstart{N}$ and $\Tblk{N}$ need to be set accordingly.
We can already notice that, using this simple first-come-first-served approach, the latest requests are highly disadvantaged if the first ones require big slices of time resources.
In the long term, as we will see in \cref{sec:results}, this could lead not only to poor performance in terms of fairness, but also to a very low admission rate.


\subsection{Max-Min Fair Scheduler} 
\label{sub:max_min_fair_scheduler}
A more flexible approach consists in dynamically adapting the duration of the allocated intervals within the admissible range, $\Tblk{n} \in [\Tmin{n}, \Tmax{n}]$, $\forall n=1,\ldots,N$, in order to distribute time resources among all traffic streams in a fairer manner.

Consider the following parameterized block duration:
\begin{equation}\label{eq:parameterized_tblk}
 \Tblk{n}(r) = \Tmin{n} + r_n(\Tmax{n} - \Tmin{n}), \quad r_n \in [0,1] \;.
\end{equation}
We consider a scheduler to be fair if $r_{\rm min} = \min_{n} \{r_n\}$ cannot be increased without breaking the limits imposed by some allocation under the \textit{strict periodicity} constraint (see \cref{sec:framework_description}).
The scheduling algorithm, then, should assign the largest possible \gls{sp} to each allocation, while respecting all the constraints.\footnote{Note that if $\Tmin{n}=\Tmax{n}$, $r_n$ has no meaning. For simplicity, this case has not been included in this study.}

To fit a new traffic stream, the pre-existing allocations will thus have to either maintain or reduce their block duration, depending on whether and how the new allocation collides with them.
This will lead to a lower rejection rate with respect to the \textit{\nameref{sub:simple_scheduler}} (\cref{sub:simple_scheduler}), and more fairness among requests distributed in time.

The proposed algorithm is here presented in two parts: the first part describes how the allocation scheme works (\cref{ssub:allocation_algorithm}), while the second part describes the fairness paradigm (\cref{ssub:optimally_fair_allocation}).

\subsubsection{Allocation Algorithm} 
\label{ssub:allocation_algorithm}

Differently from the simple scheduler, this scheduler can change the block duration within the range imposed by the requesting \gls{sta}, i.e., $\Tblk{n} \in [\Tmin{n}, \Tmax{n}]$.
To reduce the rejection rate, we check the feasibility of a new allocation $A_N$ (\cref{sub:feasibility_check_algorithm}) by assuming all existing allocations are shrunk to their minimum, i.e., $\Tblk{n} = \Tmin{n}$ for $n=1,\ldots,N$.
If $A_N$ is infeasible even under these conditions, then the allocation cannot be granted without disattending the requests of some previously accepted flow.
Therefore, $A_N$ is rejected.
Conversely, if $A_N$ is feasible, it gets accepted, and in a later step the algorithm will try to increase the resource utilization of all allocations fairly.

From now on, we use the symbol $^*$ to represent the parameter values at the end of the execution of the algorithm.
We recall that, based on the strict periodicity assumption, the starting times of the already allocated blocks cannot change


Note that, given a set of feasible allocations, reducing any $r_n$ (and, in turn, the $\Tblk{n}$) still yields a valid configuration. 
Similarly, a valid configuration for $A_N$ with $\tstart{N}$ and $r_N \geq 0$ will remain valid if $\tstart{N*} \geq \tstart{N}$ and $\tend{N*} = \tstart{N*} + \Tblk{N}(r_N^*) \leq \tend{N}$.
We thus consider the following constraints:
\begin{subequations}\label{eq:optimally_fair_alloc_constraints}
 \begin{align}
  \label{eq:rn_bound}
  r_n^*       & \leq r_n, \, \forall n \leq N;        \\
  \label{eq:tstart_bound}
  \tstart{N*} & \geq \tstart{N}; \\
  \label{eq:tend_bound}
  \tend{N*}   & \leq \tlim^N.
 \end{align}
\end{subequations}

\floatstyle{spaceruled}
\restylefloat{algorithm}
\begin{algorithm}[t!]
 \footnotesize
 \caption{Max-min fair scheduling.}
 \label{alg:complex_allocation}
 \begin{algorithmic}[1]
  \REQUIRE $A_1,\ldots, A_N$, $\Tp{1,\ldots,N}$

  \STATE Compute $\mathcal{I}_N$ considering $\Tblk{n}=\Tmin{n} \forall n=1,\ldots,N$
  \FORALL{$I_m^N = \qty[\tfeas, \tlim]^N_m \in \mathcal{I}_N$}
  \STATE $\tstart{N} \leftarrow \tfeas$
  \STATE Set $r_N$ such that $\Tblk{N} = \min\qty{\Tmax{N}, \tlim - \tfeas}$ \COMMENT{\cref{eq:parameterized_tblk}}

  \FORALL{$A_n, \; n=1,\ldots,N-1$}

  \FORALL{(block $k \in A_n$) $\in \Tp{1,\ldots,N}$}
  \IF{block $k$ collides with $A_N$}
  \STATE Update $r_n^*$, $r_N^*$, $\tstart{N*}$ \COMMENT{\cref{ssub:optimally_fair_allocation}}

  \IF{$r_n^* < r_n$}
  \STATE Add/update $A_n$ to a list $\mathcal{C}$ of colliding allocations
  \STATE Memorize $r_{n,\rm prev} \leftarrow r_n$
  \ENDIF

  \STATE Update $r_n$, $r_N$, $\tstart{N}$
  \ENDIF

  \ENDFOR
  \ENDFOR

  \FORALL{$A_n \in \mathcal{C}$}
  \STATE Compute $\Delta t= \tlim - \tstart{n}$ for $A_n$ given $A_N$ \COMMENT{see \cref{sec:scheduling_algorithms}}
  \STATE $\Tblk{n} \leftarrow \min\qty{\Tblk{n}(r_{n,\rm prev}), \Delta t}$ \COMMENT{Try to improve the allocation duration if $A_N$ has been further reduced}
  \ENDFOR

  \STATE Compute allocation score $s_m \leftarrow \min_{n=1,\ldots,N} r_n$
  \ENDFOR

  \RETURN The configuration which maximizes the allocation score $\qty{s_m}$
 \end{algorithmic}
\end{algorithm}

The algorithm starts by considering the first feasible interval $I_1^N$, which ensures a valid configuration when $r_n=0$, $\forall n \leq N$.
The new request is temporarily accepted with $\tstart{N}=\tfeasmn{1}{N}$ and maximum possible $r_N$, such that $\Tblk{N} = \min\qty{\Tmax{N}, \tlim - \tfeas}$.

Then, the algorithms try to re-balance the resource allocation by increasing all $\{ r_n, \, \forall n \leq N \}$ to their previous values.
Given that feasible intervals $\mathcal{I}_N$ were computed considering all allocations with minimum duration, though, setting $\tstart{N}=\tfeasmn{1}{N}$ may (or may not) create a collision with a generic $A_i$ when setting $r_i\geq 0$ back to its previous value.

On the other hand, thanks to the information given by $\tlim$, we can always choose $r_N$ such that the new allocation does not collide with a previous one, on the right.

Collisions can be found iteratively over each block of each previous allocation in a joint period.

\begin{figure}[tbp]
  \centering
  \newcommand\fscale{1}
  \begin{tikzpicture}[scale=\fscale, every node/.style={transform shape, font=\footnotesize}]

\definecolor{color0}{rgb}{0.1725,0.4824,0.7137}
\definecolor{color1}{rgb}{0.9922,0.6824,0.3804}

\def\allocHeight{.7}

\draw[thick, dashed] (-1,0) -- (-.5,0);
\draw[thick] (-.5,0) -- (4.5,0);
\draw[thick, dashed] (4.5,0) -- (5,0);

\draw[fill=color0] (0,0) rectangle ++(2,\allocHeight) node {};
\draw[fill=color1] (2,0) rectangle ++(1.7,\allocHeight) node {};

\draw[pattern=north west lines] (0,0) rectangle ++(1.2,\allocHeight) node {};
\draw[pattern=north west lines] (2,0) rectangle ++(1.2,\allocHeight) node {};

\draw[pattern color=color0, pattern=vertical lines] (2,0) rectangle (2.5,\allocHeight) node {};

\draw[|-|] (0,-0.2) -- node[below] {$\Tmin{n}$} ++(1.2, 0);
\draw[|-|] (0,\allocHeight+0.2) -- node[above] {$\Tblk{n}(r_n)$} ++(2.5, 0);

\draw[|-|] (2,\allocHeight+0.5) -- node[above] {$\Tmin{N}$} ++(1.2, 0);
\draw[|-|] (2,-0.2) -- node[below] {$\Tblk{N}(r_N)$} ++(1.7, 0);

\draw (0,0) -- ++(0,-.4) node[below] {$t_{k, \rm start}^{n}$};
\draw (2,0) -- ++(0,-.4) node[below] {$t_{h, \rm start}^{N}$};

\draw (4,\allocHeight+0.4) -- (4,-.4) node[below right] {$\tlim$};

\end{tikzpicture}
  \caption{Representation of a collision between $A_n$ and $A_N$.}
  \label{fig:collision_baseline}
 \end{figure}
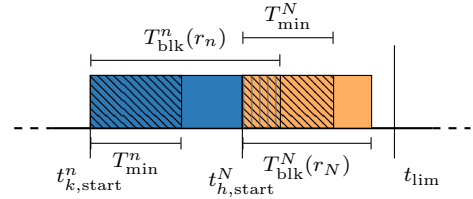

If for a certain block $b_k^n \in A_n$ and a block $b_h^N \in A_N$ we have
\begin{equation}
 t_{k, \rm start}^{n} + \Tblk{n}(r_n) \geq t_{h, \rm start}^{N}.
\end{equation}
then the two allocations are in conflict, as shown in \cref{fig:collision_baseline}.
In this case, $\tstart{N*}$, $r_N^*$, and $r_n^*$ have to be updated following the constraints in \eqref{eq:optimally_fair_alloc_constraints}, as described in \cref{ssub:optimally_fair_allocation}.

The constraints from \eqref{eq:optimally_fair_alloc_constraints}, the existence of a non-empty set of feasible intervals, and the iterative nature of the problem ensure that the algorithm will stop in a finite time with a valid configuration.
Since each feasible interval $I_m^N \in \mathcal{I}_N$ has one locally fairest configuration, the exhaustive search described in \cref{alg:complex_allocation} is able to find the globally fairest configuration by exhaustive search.


\subsubsection{Optimally fair allocation} 
\label{ssub:optimally_fair_allocation}

In this section, we will discuss how fairness can be achieved given a pair of colliding allocations $A_n$, $n\in \{1,\ldots,N-1\}$, and $A_N$.
In \cref{ssub:allocation_algorithm} we explained how such a collision can be found, e.g, between blocks $b_k^n$ and $b_h^N$.
For the sake of clarity, in this section we will drop the notation for the specific colliding blocks. 

In order to fully exploit the available resources, looking at \cref{fig:collision_baseline}, we force $\tstart{N} = \tend{n}$ and $\tend{N} \leq \tlim$.
In this way we make $A_N$ start right after $A_n$, still respecting the limits imposed by $\tlim$.

While possibly not being optimal, this is still a sensible choice for a greedy approach that tries to maximize the fairness of the current configuration.
By doing so and imposing $r_n = r_N = r^*$, what we call the \textit{fairness equation}, and by noting that $r^*\leq 1$ should hold, we have that
\begin{equation}\label{eq:bounded_rstar}
 r^* = \min \qty{1, \frac{\tlim - \tstart{n} - \Tmin{n} - \Tmin{N}}{\qty(\Tmax{n} - \Tmin{n}) + \qty(\Tmax{N} - \Tmin{N})}}.
\end{equation}
We call $r^*$ the \textit{fair allocation ratio}, and note that if $r^*<1$, it must be that $\tend{N}=\tlim$, whereas if $r^*=1$ in general $\tend{N} \leq \tlim$ by construction.

Depending on the initial conditions of the problem, there is a number of different cases which have to be properly managed in order to obtain a fair distribution of resources.

First of all, if $r_n \leq r^*$, following \cref{eq:rn_bound}, it means that previous adjustments do not make it possible for $A_n$ to obtain more resources while still ensuring a valid configuration, and thus $r_n^* = r_n$.
Furthermore, since we assume that a collision happens between $A_n$ and $A_N$ with this configuration, $A_N$ has to be delayed setting $\tstart{N*} = \tstart{n} + \Tblk{n}(r_n^*) > \tstart{N}$.

In case also $r_N \leq r^*$, allocation $A_N$ cannot be extended either.
Since both allocations have $r_n, r_N \leq r^*$, they will both surely fit in the feasible interval.
If, instead, $r_N > r^*$, $A_N$ can obtain $r_N^* \geq r_n^*$, i.e., $\Tblk{N*} = \min\qty{\Tblk{N}(r_N), \tlim - \tstart{N*}}$.

On the other hand, if $1 \geq r_n > r^*$, the block duration must be reduced so that $r_n^* = r^*$.
Then, if also $r_N > r^*$, both allocations must be trimmed and are fairly allocated, i.e., $r_n^*=r_N^*=r^* < 1$.
It follows from the properties of \cref{eq:bounded_rstar} that $\tend{N*} = \tlim$ and $\tstart{N*} > \tstart{N}$.

Finally, if $r_N \leq r^* < r_n \leq 1$, and therefore $r_N < 1$, the properties of \cref{eq:bounded_rstar} imply that $\tend{N}=\tlim$.
Since $A_N$ cannot be extended without possibly reducing the allocation ratio of other allocations, $\tstart{N*}=\tstart{N}$ and $r_N^* = r_N$.
Since, by assumption, $\tend{n} > \tstart{N}$, the duration of $A_n$ needs to be reduced so that $\Tblk{n*} = \tstart{N*} - \tstart{n} < \Tblk{n}$.



\section{Results} 
\label{sec:results}

\begin{figure*}[t!]
\newcommand\fheight{0.5\columnwidth}
\newcommand\fwidth{0.9\columnwidth}

 \begin{subfigure}[b]{\textwidth}
  \centering
\begin{tikzpicture}

\definecolor{color0}{rgb}{0.1725,0.4824,0.7137}
\definecolor{color1}{rgb}{0.9922,0.6824,0.3804}

\pgfplotsset{every tick label/.append style={font=\scriptsize}}
\tikzstyle{every node}=[font=\scriptsize]

\begin{axis}[
width=0,
height=0,
at={(0,0)},
scale only axis,
xmin=0,
xmax=0,
xtick={},
ymin=0,
ymax=0,
ytick={},
axis background/.style={fill=white},
legend style={at={(0.5,0.5)}, anchor=center, draw=white!80.0!black,
              /tikz/every even column/.append style={column sep=4em}},
legend columns=2
]
\addplot [line width=1.5pt, color0]
table {%
0 0
};
\addlegendentry{Simple}
\addplot [line width=1.5pt, color1]
table {%
0 0
};
\addlegendentry{Max-Min Fair}
\end{axis}

\end{tikzpicture}
 \end{subfigure}
 \\
 \vspace{-2ex}
 \\
 \hspace*{\fill}%
 \begin{subfigure}[t]{0.45\textwidth}
  \centering
\begin{tikzpicture}

\definecolor{color0}{rgb}{0.1725,0.4824,0.7137}
\definecolor{color1}{rgb}{0.9922,0.6824,0.3804}

\pgfplotsset{every tick label/.append style={font=\scriptsize}}
\tikzstyle{every node}=[font=\scriptsize]

\begin{axis}[
width=\fwidth,
height=\fheight,
at={(0,0)},
legend cell align={left},
legend style={at={(0.97,0.03)}, anchor=south east, draw=white!80.0!black},
tick align=outside,
tick pos=left,
x grid style={white!69.01960784313725!black},
xlabel={$\rho$},
xmin=0, xmax=1,
xtick style={color=black},
y grid style={white!69.01960784313725!black},
ylabel={Acceptance Rate},
ymin=-0.05, ymax=1.05,
ytick style={color=black},
ymajorgrids,
xmajorgrids,
xlabel style={font=\scriptsize\color{white!15!black}},
ylabel style={font=\scriptsize\color{white!15!black}}
]
\addplot [color0, line width=1.5pt]
table {%
0.01 0.06
0.03 0.06
0.05 0.06
0.07 0.0789473684210526
0.09 0.1
0.11 0.12
0.13 0.13953488372093
0.15 0.157894736842105
0.17 0.176470588235294
0.19 0.193548387096774
0.21 0.214285714285714
0.23 0.230769230769231
0.25 0.24
0.27 0.304347826086957
0.29 0.318181818181818
0.31 0.333333333333333
0.33 0.35
0.35 0.368421052631579
0.37 0.388888888888889
0.39 0.411764705882353
0.41 0.411764705882353
0.43 0.4375
0.45 0.4375
0.47 0.466666666666667
0.49 0.466666666666667
0.51 0.571428571428571
0.53 0.571428571428571
0.55 0.571428571428571
0.57 0.615384615384615
0.59 0.615384615384615
0.61 0.615384615384615
0.63 0.666666666666667
0.65 0.666666666666667
0.67 0.666666666666667
0.69 0.666666666666667
0.71 0.666666666666667
0.73 0.727272727272727
0.75 0.727272727272727
0.77 0.818181818181818
0.79 0.818181818181818
0.81 0.818181818181818
0.83 0.818181818181818
0.85 0.9
0.87 0.9
0.89 0.9
0.91 0.9
0.93 0.9
0.95 0.9
0.97 0.9
0.99 0.9
};
\addplot [color1, line width=1.5pt]
table {%
0.01 1
0.03 1
0.05 0.8
0.07 0.631578947368421
0.09 0.8
0.11 0.8
0.13 0.558139534883721
0.15 0.631578947368421
0.17 0.705882352941177
0.19 0.774193548387097
0.21 0.857142857142857
0.23 0.923076923076923
0.25 0.76
0.27 0.608695652173913
0.29 0.636363636363636
0.31 0.666666666666667
0.33 0.7
0.35 0.736842105263158
0.37 0.777777777777778
0.39 0.823529411764706
0.41 0.823529411764706
0.43 0.875
0.45 0.875
0.47 0.933333333333333
0.49 0.933333333333333
0.51 0.571428571428571
0.53 0.571428571428571
0.55 0.571428571428571
0.57 0.615384615384615
0.59 0.615384615384615
0.61 0.615384615384615
0.63 0.666666666666667
0.65 0.666666666666667
0.67 0.75
0.69 0.75
0.71 0.75
0.73 0.818181818181818
0.75 0.818181818181818
0.77 0.818181818181818
0.79 0.818181818181818
0.81 0.818181818181818
0.83 0.818181818181818
0.85 0.9
0.87 0.9
0.89 0.9
0.91 0.9
0.93 0.9
0.95 0.9
0.97 0.9
0.99 0.9
};
\end{axis}

\begin{axis}[
width=\fwidth,
height=\fheight,
at={(0,0)},
legend cell align={left},
legend style={font=\scritpsize,at={(0.03,0.97)}, anchor=north west, draw=white!80.0!black},
xmin=0, xmax=1,
xtick={\empty},
ylabel={$N_{\rm max}$},
ylabel style={font=\scriptsize\color{red}},
ytick align=outside,
yticklabel pos=right,
ytick pos=right,
y tick style={color=red},
every y tick label/.append style={red},
ymin=-5, ymax=105,
]
\addplot [thick, red, dashed]
table {%
0.01 100
0.03 100
0.05 100
0.07 76
0.09 60
0.11 50
0.13 43
0.15 38
0.17 34
0.19 31
0.21 28
0.23 26
0.25 25
0.27 23
0.29 22
0.31 21
0.33 20
0.35 19
0.37 18
0.39 17
0.41 17
0.43 16
0.45 16
0.47 15
0.49 15
0.51 14
0.53 14
0.55 14
0.57 13
0.59 13
0.61 13
0.63 12
0.65 12
0.67 12
0.69 12
0.71 12
0.73 11
0.75 11
0.77 11
0.79 11
0.81 11
0.83 11
0.85 10
0.87 10
0.89 10
0.91 10
0.93 10
0.95 10
0.97 10
0.99 10
};
\end{axis}

\end{tikzpicture}
  \vspace*{-2ex}
  \caption{Acceptance Rate vs. $\rho$.}
  \label{fig:accrate_rho_bound}
 \end{subfigure}
 \hspace*{\fill}%
 \begin{subfigure}[t]{0.45\textwidth}
  \centering
\begin{tikzpicture}

\definecolor{color0}{rgb}{0.1725,0.4824,0.7137}
\definecolor{color1}{rgb}{0.9922,0.6824,0.3804}

\pgfplotsset{every tick label/.append style={font=\scriptsize}}
\tikzstyle{every node}=[font=\scriptsize]

\begin{axis}[
width=\fwidth,
height=\fheight,
at={(0,0)},
legend cell align={left},
legend style={at={(0.97,0.03)}, anchor=south east, draw=white!80.0!black},
tick align=outside,
tick pos=left,
x grid style={white!69.01960784313725!black},
xlabel={$\rho$},
xmajorgrids,
xmin=0, xmax=1,
xtick style={color=black},
y grid style={white!69.01960784313725!black},
ylabel={$\mathcal{J}(\Tblk{})$},
ymajorgrids,
ymin=0.841528500038187, ymax=1.00754626190294,
ytick style={color=black},
xlabel style={font=\scriptsize\color{white!15!black}},
ylabel style={font=\scriptsize\color{white!15!black}}
]
\addplot [line width=1.5pt, color0]
table {%
0.01 0.849319465179908
0.03 0.880004311078521
0.05 0.907245991749911
0.07 0.931191025373444
0.09 0.951432469736029
0.11 0.968058170694157
0.13 0.981053255477509
0.15 0.990583185862603
0.17 0.996688478464585
0.19 0.999639505950568
0.21 1
0.23 1
0.25 1
0.27 0.940785732255936
0.29 0.958069515803483
0.31 0.972374834768233
0.33 0.983519710216846
0.35 0.991812906296318
0.37 0.997113113867553
0.39 0.999683658340229
0.41 1
0.43 1
0.45 1
0.47 1
0.49 1
0.51 0.97566779919645
0.53 0.98555129992064
0.55 0.992740992989523
0.57 0.997448595942913
0.59 0.999714079780002
0.61 1
0.63 1
0.65 1
0.67 1
0.69 1
0.71 1
0.73 1
0.75 1
0.77 0.9976565533181
0.79 0.999731712840428
0.81 1
0.83 1
0.85 1
0.87 1
0.89 1
0.91 1
0.93 1
0.95 1
0.97 1
0.99 1
};
\addplot [line width=1.5pt, color1]
table {%
0.01 0.921868435555719
0.03 0.935607420849374
0.05 0.991059971660837
0.07 0.971273866299012
0.09 0.979975176689939
0.11 0.962314993105445
0.13 0.992325957740727
0.15 0.996206115548616
0.17 0.998669187816993
0.19 0.999854341774227
0.21 0.999676320266437
0.23 0.997068376261785
0.25 0.90760609857301
0.27 0.974428395252188
0.29 0.982077436358595
0.31 0.988290022547726
0.33 0.993058135685889
0.35 0.996566540461579
0.37 0.998791672904117
0.39 0.999867000855054
0.41 0.999706662978072
0.43 0.997396622648628
0.45 0.992989802431929
0.47 0.986623071493426
0.49 0.978614174142177
0.51 0.989413709772276
0.53 0.993747620960595
0.55 0.996870051452327
0.57 0.998901427265443
0.59 0.999876267577453
0.61 1
0.63 1
0.65 1
0.67 0.979183205152572
0.69 0.98532077537319
0.71 0.990330321005357
0.73 0.994245883897733
0.75 0.997109409570821
0.77 0.998969715584004
0.79 0.99988136380775
0.81 1
0.83 1
0.85 1
0.87 1
0.89 1
0.91 1
0.93 1
0.95 1
0.97 1
0.99 1
};
\end{axis}

\end{tikzpicture}
  \caption{Jain's Index vs. $\rho$.}
  \label{fig:jain_rho}
 \end{subfigure}
 \hspace*{\fill}%
 \\
 \vspace*{2ex}
 \\
 \hspace*{-2pt}
 \begin{subfigure}[t]{0.45\textwidth}
  \centering
\begin{tikzpicture}

\definecolor{color0}{rgb}{0.1725,0.4824,0.7137}
\definecolor{color1}{rgb}{0.9922,0.6824,0.3804}

\pgfplotsset{every tick label/.append style={font=\scriptsize}}
\tikzstyle{every node}=[font=\scriptsize]

\begin{axis}[
width=\fwidth,
height=\fheight,
at={(0,0)},
legend cell align={left},
legend style={at={(0.03,0.97)}, anchor=north west, draw=white!80.0!black},
tick align=outside,
tick pos=left,
x grid style={white!69.01960784313725!black},
xlabel={$\rho$},
xmajorgrids,
xmin=0, xmax=1,
xtick style={color=black},
y grid style={white!69.01960784313725!black},
ylabel={$\overline{\Tblk{}}/\Tmax{}$},
ymajorgrids,
ymin=0, ymax=1.03,
ytick style={color=black},
xlabel style={font=\scriptsize\color{white!15!black}},
ylabel style={font=\scriptsize\color{white!15!black}},
legend style={font=\scriptsize}
]
\addplot [line width=1.5pt, color0]
table {%
0.01 0.841489898989899
0.03 0.858265080100963
0.05 0.874894980046209
0.07 0.891609140043881
0.09 0.908231125647315
0.11 0.924868165417707
0.13 0.941487257727298
0.15 0.958218311479181
0.17 0.974869814522263
0.19 0.991578884722966
0.21 1
0.23 1
0.25 1
0.27 0.907093858885017
0.29 0.921313832287318
0.31 0.935618823000898
0.33 0.949806179455022
0.35 0.964235076246419
0.37 0.978505358978124
0.39 0.992790108744228
0.41 1
0.43 1
0.45 1
0.47 1
0.49 1
0.51 0.943673538740372
0.53 0.956238521579431
0.55 0.968691860465116
0.57 0.981242638398115
0.59 0.993648616412214
0.61 1
0.63 1
0.65 1
0.67 1
0.69 1
0.71 1
0.73 1
0.75 1
0.77 0.983153360476781
0.79 0.994241556271631
0.81 1
0.83 1
0.85 1
0.87 1
0.89 1
0.91 1
0.93 1
0.95 1
0.97 1
0.99 1
};
\addplot [line width=1.5pt, color1]
table {%
0.01 0.0504878787878788
0.03 0.0514943594498531
0.05 0.065615154379332
0.07 0.111447797934393
0.09 0.113525483783047
0.11 0.138726061615321
0.13 0.235364751087755
0.15 0.239547389003911
0.17 0.243710139839682
0.19 0.24788728203297
0.21 0.252026986143886
0.23 0.256212708371778
0.25 0.328917755754728
0.27 0.453533318815331
0.29 0.460643092150737
0.31 0.467795372866127
0.33 0.474888838216851
0.35 0.482103070111982
0.37 0.489237997357216
0.39 0.496380157902577
0.41 0.503520807519115
0.43 0.510650688080424
0.45 0.517760029833121
0.47 0.524954313441301
0.49 0.532113314809492
0.51 0.943645219755324
0.53 0.956209825528007
0.55 0.968662790697674
0.57 0.981213191990577
0.59 0.993618797709924
0.61 1
0.63 1
0.65 1
0.67 0.927698432584426
0.69 0.938753662384494
0.71 0.949831822587082
0.73 0.960925108861782
0.75 0.972025320148187
0.77 0.983123856729805
0.79 0.994211719775629
0.81 1
0.83 1
0.85 1
0.87 1
0.89 1
0.91 1
0.93 1
0.95 1
0.97 1
0.99 1
};
\end{axis}

\end{tikzpicture}
  \caption{Normalized average $\Tblk{}$ vs. $\rho$.}
  \label{fig:tblk_rho}
 \end{subfigure}
 \hspace*{6ex}%
 \begin{subfigure}[t]{0.45\textwidth}
  \centering
\begin{tikzpicture}

\definecolor{color0}{rgb}{0.1725,0.4824,0.7137}
\definecolor{color1}{rgb}{0.9922,0.6824,0.3804}

\pgfplotsset{every tick label/.append style={font=\scriptsize}}
\tikzstyle{every node}=[font=\scriptsize]

\begin{axis}[
width=\fwidth,
height=\fheight,
at={(0,0)},
legend cell align={left},
legend style={at={(0.03,0.03)}, anchor=south west, font=\scriptsize, draw=white!80.0!black},
tick align=outside,
tick pos=left,
x grid style={white!69.01960784313725!black},
xlabel={Acceptance Rate},
xmajorgrids,
xmin=0, xmax=1.03,
xtick style={color=black},
y grid style={white!69.01960784313725!black},
ylabel={$\overline{\Tblk{}}/\Tmax{}$},
ymajorgrids,
ymin=0, ymax=1.03,
ytick style={color=black},
xlabel style={font=\scriptsize\color{white!15!black}},
ylabel style={font=\scriptsize\color{white!15!black}}
]
\addplot [thick, color0, mark=*, mark size=2, mark options={solid,fill opacity=0}, only marks]
table {%
0.06 0.841489898989899
0.06 0.858265080100963
0.06 0.874894980046209
0.0789473684210526 0.891609140043881
0.1 0.908231125647315
0.12 0.924868165417707
0.13953488372093 0.941487257727298
0.157894736842105 0.958218311479181
0.176470588235294 0.974869814522263
0.193548387096774 0.991578884722966
0.214285714285714 1
0.230769230769231 1
0.24 1
0.304347826086957 0.907093858885017
0.318181818181818 0.921313832287318
0.333333333333333 0.935618823000898
0.35 0.949806179455022
0.368421052631579 0.964235076246419
0.388888888888889 0.978505358978124
0.411764705882353 0.992790108744228
0.411764705882353 1
0.4375 1
0.4375 1
0.466666666666667 1
0.466666666666667 1
0.571428571428571 0.943673538740372
0.571428571428571 0.956238521579431
0.571428571428571 0.968691860465116
0.615384615384615 0.981242638398115
0.615384615384615 0.993648616412214
0.615384615384615 1
0.666666666666667 1
0.666666666666667 1
0.666666666666667 1
0.666666666666667 1
0.666666666666667 1
0.727272727272727 1
0.727272727272727 1
0.818181818181818 0.983153360476781
0.818181818181818 0.994241556271631
0.818181818181818 1
0.818181818181818 1
0.9 1
0.9 1
0.9 1
0.9 1
0.9 1
0.9 1
0.9 1
0.9 1
};
\addlegendentry{Simple}
\addplot [thick, color1, mark=x, mark size=2, mark options={solid}, only marks]
table {%
1 0.0504878787878788
1 0.0514943594498531
0.8 0.065615154379332
0.631578947368421 0.111447797934393
0.8 0.113525483783047
0.8 0.138726061615321
0.558139534883721 0.235364751087755
0.631578947368421 0.239547389003911
0.705882352941177 0.243710139839682
0.774193548387097 0.24788728203297
0.857142857142857 0.252026986143886
0.923076923076923 0.256212708371778
0.76 0.328917755754728
0.608695652173913 0.453533318815331
0.636363636363636 0.460643092150737
0.666666666666667 0.467795372866127
0.7 0.474888838216851
0.736842105263158 0.482103070111982
0.777777777777778 0.489237997357216
0.823529411764706 0.496380157902577
0.823529411764706 0.503520807519115
0.875 0.510650688080424
0.875 0.517760029833121
0.933333333333333 0.524954313441301
0.933333333333333 0.532113314809492
0.571428571428571 0.943645219755324
0.571428571428571 0.956209825528007
0.571428571428571 0.968662790697674
0.615384615384615 0.981213191990577
0.615384615384615 0.993618797709924
0.615384615384615 1
0.666666666666667 1
0.666666666666667 1
0.75 0.927698432584426
0.75 0.938753662384494
0.75 0.949831822587082
0.818181818181818 0.960925108861782
0.818181818181818 0.972025320148187
0.818181818181818 0.983123856729805
0.818181818181818 0.994211719775629
0.818181818181818 1
0.818181818181818 1
0.9 1
0.9 1
0.9 1
0.9 1
0.9 1
0.9 1
0.9 1
0.9 1
};
\addlegendentry{Max-Min Fair}
\end{axis}

\end{tikzpicture}
  \caption{Normalized average $\Tblk{}$ vs. Acceptance Rate.}
  \label{fig:tblk_accrate}
 \end{subfigure}
 \hspace*{\fill}%

 \caption{Results for Scenario~1.}
 \label{fig:results_scenario_1}
\end{figure*}
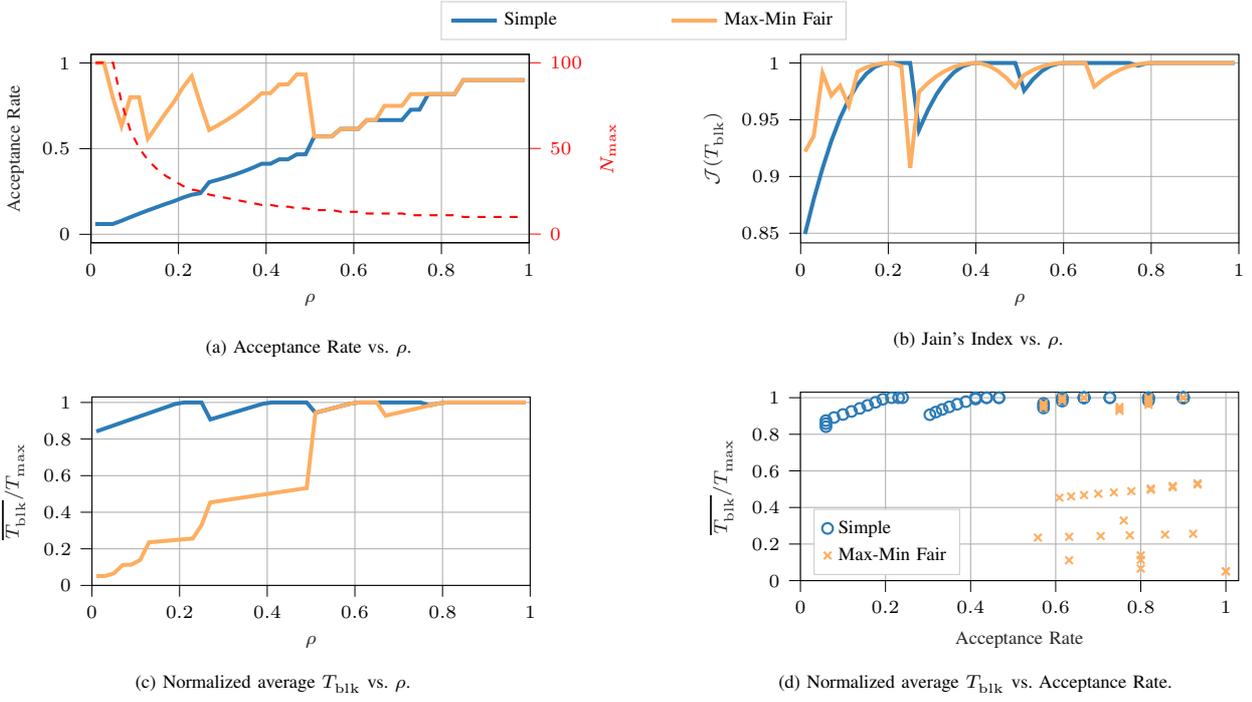

In this section, we evaluate the algorithms described in \cref{sec:scheduling_algorithms}.
The proposed schedulers have been implemented in Python, only focusing on their capabilities of allocating communication resources to the different traffic streams.
The simulations proposed here do not include full stack behaviors, which will be investigated in future works, and their aim is thus to highlight the fundamental characteristics of each algorithm.

Based on their design criteria, we expected the two algorithms to differ mainly with respect to three \glspl{kpi}, namely the acceptance rate of new requests, the fairness among accepted allocations, and the average scheduled block duration.
To highlight these characteristics we shape the offered traffic based on three parameters, namely: the \textit{average allocation request}
\begin{equation}\label{eq:avg_allocation_request}
T_{\rm avg} = \frac{\Tmin{} + \Tmax{}}{2};
\end{equation}
the \textit{interval ratio}
\begin{equation}\label{eq:interval_ratio}
\rho = \frac{\Tmin{}}{\Tmax{}}  \in [0, 1];
\end{equation}
and the \textit{load factor}
\begin{equation}\label{eq:load_factor}
\lambda = \frac{T_{\rm avg}}{\Tp{}}.
\end{equation}
Note that, for a given average allocation request $T_{\rm avg}$, a low interval ratio $\rho$ corresponds to very flexible allocations, while $\rho=1$ corresponds to rigid allocations where $\Tmin{}=\Tmax{}$.

The proposed algorithms are compared in two different simulation scenarios:
\begin{itemize}
 \item \textit{Scenario~1}: all traffic streams are homogeneous, i.e., all requests have the same parameters.
       Specifically, we consider the case with periodicity $\Tp{}=\frac{\TBI}{3}$, load factor $\lambda=0.1$, and $\rho \in (0,1)$.
       The impact of different periodicities and load factors is also discussed.
 \item \textit{Scenario~2}: multiple non-homogeneous applications coexist on the same network, thus generating traffic streams with different characteristics.
      We analyze a scenario where traffic streams can be of class $C_1$ or $C_2$, with periodicity $\Tp{C_1}=\frac{\TBI}{3}$ and $\Tp{C_2}=\frac{\TBI}{5}$, respectively.
       Both classes have load factor $\lambda=0.1$ and interval ratio $\rho=0.1$.
\end{itemize}

To evaluate the performance of the algorithms, we propose three \glspl{kpi} for \textit{Scenario~1}, shown in \cref{fig:results_scenario_1}.
Note that, given the discrete behavior of the problem and the lack of randomness in the proposed algorithms, the plots cannot be smoothed by running multiple repetitions.

\begin{figure*}[t!]
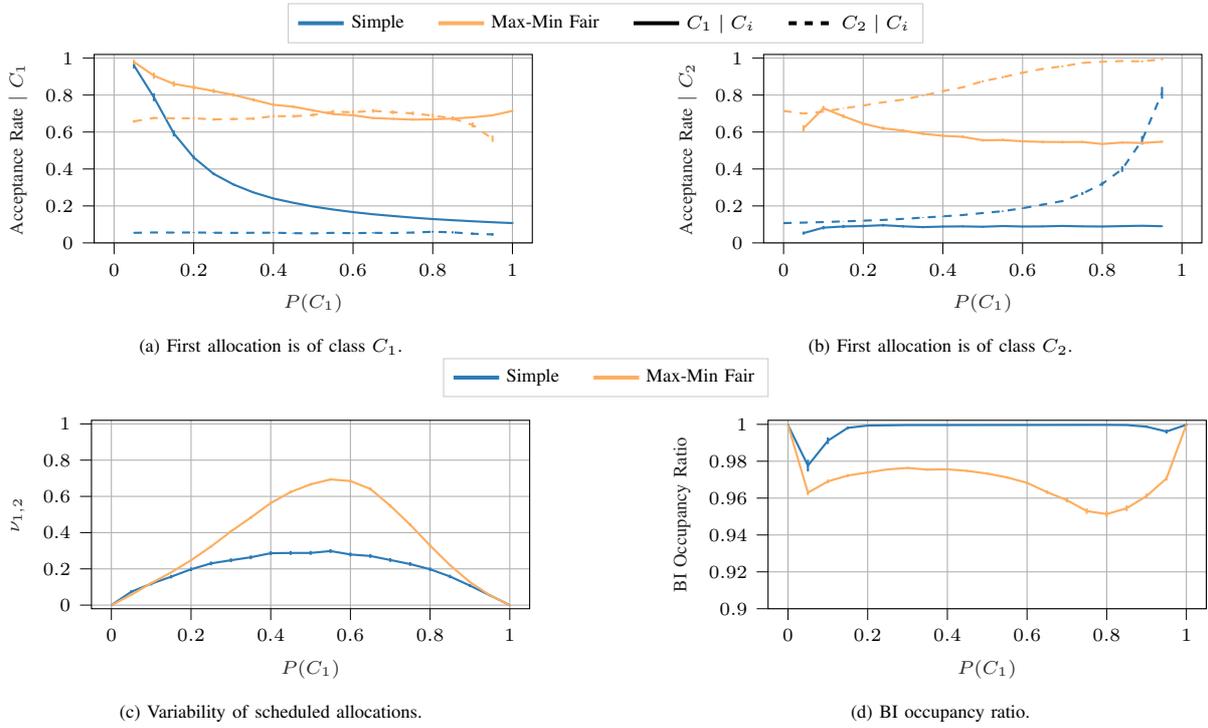

 \newcommand\fheight{0.5\columnwidth}
 \newcommand\fwidth{0.9\columnwidth}
 \begin{subfigure}[b]{\textwidth}
  \centering
\begin{tikzpicture}

\definecolor{color0}{rgb}{0.1725,0.4824,0.7137}
\definecolor{color1}{rgb}{0.9922,0.6824,0.3804}
\definecolor{color2}{rgb}{0.172549019607843,0.627450980392157,0.172549019607843}
\definecolor{color3}{rgb}{0.83921568627451,0.152941176470588,0.156862745098039}

\pgfplotsset{every tick label/.append style={font=\scriptsize}}
\tikzstyle{every node}=[font=\scriptsize]

\begin{axis}[
width=0,
height=0,
at={(0,0)},
scale only axis,
xmin=0,
xmax=0,
xtick={},
ymin=0,
ymax=0,
ytick={},
axis background/.style={fill=white},
legend style={at={(0.5,0.5)}, anchor=center, draw=white!80.0!black,
              /tikz/every even column/.append style={column sep=1em}},
legend columns=4
]
\addplot [line width=1.5pt, color0]
table {%
-10 -10
};
\addlegendentry{Simple}
\addplot [line width=1.5pt, color1]
table {%
-10 -10
};
\addlegendentry{Max-Min Fair}

\addplot [line width=1.5pt, black]
table {%
-10 -10
};
\addlegendentry{$C_1 \; | \; C_i$}
\addplot [line width=1.5pt, black, dashed]
table {%
-10 -10
};
\addlegendentry{$C_2 \; | \; C_i$}
\end{axis}

\end{tikzpicture}
 \end{subfigure}
 \\
 \hspace*{\fill}%
 \begin{subfigure}[b]{0.45\textwidth}
  \centering
  \input{img/conditional_acceptance_c1_3000run_21pt.tex}
  \caption{First allocation is of class $C_1$.}
  \label{fig:conditional_c1}
 \end{subfigure}
 \hspace*{\fill}%
 \begin{subfigure}[b]{0.45\textwidth}
  \centering
  \input{img/conditional_acceptance_c2_3000run_21pt.tex}
  \caption{First allocation is of class $C_2$.}
  \label{fig:conditional_c2}
 \end{subfigure}
 \hspace*{\fill}%
 \\
 \vspace{1ex}%
 \begin{subfigure}[b]{\textwidth}
  \centering
\begin{tikzpicture}

\definecolor{color0}{rgb}{0.1725,0.4824,0.7137}
\definecolor{color1}{rgb}{0.9922,0.6824,0.3804}
\definecolor{color2}{rgb}{0.172549019607843,0.627450980392157,0.172549019607843}
\definecolor{color3}{rgb}{0.83921568627451,0.152941176470588,0.156862745098039}

\pgfplotsset{every tick label/.append style={font=\scriptsize}}
\tikzstyle{every node}=[font=\scriptsize]

\begin{axis}[
width=0,
height=0,
at={(0,0)},
scale only axis,
xmin=0,
xmax=0,
xtick={},
ymin=0,
ymax=0,
ytick={},
axis background/.style={fill=white},
legend style={at={(0.5,0.5)}, anchor=center, draw=white!80.0!black,
              /tikz/every even column/.append style={column sep=1em}},
legend columns=4
]
\addplot [line width=1.5pt, color0]
table {%
-10 -10
};
\addlegendentry{Simple}
\addplot [line width=1.5pt, color1]
table {%
-10 -10
};
\addlegendentry{Max-Min Fair}

\end{axis}

\end{tikzpicture}
 \end{subfigure}
 \\
 \hspace*{\fill}%
 \begin{subfigure}[b]{0.45\textwidth}
  \centering
\begin{tikzpicture}

\definecolor{color0}{rgb}{0.1725,0.4824,0.7137}
\definecolor{color1}{rgb}{0.9922,0.6824,0.3804}

\pgfplotsset{every tick label/.append style={font=\scriptsize}}
\tikzstyle{every node}=[font=\scriptsize]

\begin{axis}[
width=\fwidth,
height=\fheight,
at={(0,0)},
legend cell align={left},
legend style={font=\scriptsize, at={(0.03,0.03)}, anchor=south west, draw=white!80!black,
              /tikz/every even column/.append style={column sep=1em}},
tick align=outside,
tick pos=left,
x grid style={white!69.01960784313725!black},
xlabel={$P(C_1)$},
xmajorgrids,
xmin=-0.05, xmax=1.05,
xtick style={color=black},
y grid style={white!69.01960784313725!black},
ylabel={$\nu_{1,2}$},
ymajorgrids,
ymin=-0.02, ymax=1.02,
ytick style={color=black},
xlabel style={font=\scriptsize\color{white!15!black}},
ylabel style={font=\scriptsize\color{white!15!black}},
]
\path [draw=color0, thick]
(axis cs:0,0)
--(axis cs:0,0);

\path [draw=color0, thick]
(axis cs:0.05,0.0655630616222442)
--(axis cs:0.05,0.0820258272666447);

\path [draw=color0, thick]
(axis cs:0.1,0.108825709495144)
--(axis cs:0.1,0.126974290504856);

\path [draw=color0, thick]
(axis cs:0.15,0.147235497715346)
--(axis cs:0.15,0.166264502284654);

\path [draw=color0, thick]
(axis cs:0.2,0.187576044967238)
--(axis cs:0.2,0.207790621699429);

\path [draw=color0, thick]
(axis cs:0.25,0.219732258344294)
--(axis cs:0.25,0.240634408322372);

\path [draw=color0, thick]
(axis cs:0.3,0.237029665909033)
--(axis cs:0.3,0.258170334090967);

\path [draw=color0, thick]
(axis cs:0.35,0.253342212762491)
--(axis cs:0.35,0.274657787237509);

\path [draw=color0, thick]
(axis cs:0.4,0.275875352339836)
--(axis cs:0.4,0.297324647660164);

\path [draw=color0, thick]
(axis cs:0.45,0.277073518497724)
--(axis cs:0.45,0.298526481502276);

\path [draw=color0, thick]
(axis cs:0.5,0.277273229600667)
--(axis cs:0.5,0.298726770399333);

\path [draw=color0, thick]
(axis cs:0.55,0.28766479055469)
--(axis cs:0.55,0.30913520944531);

\path [draw=color0, thick]
(axis cs:0.6,0.268689976990003)
--(axis cs:0.6,0.290110023009997);

\path [draw=color0, thick]
(axis cs:0.65,0.260514220925361)
--(axis cs:0.65,0.281885779074639);

\path [draw=color0, thick]
(axis cs:0.7,0.238420901134785)
--(axis cs:0.7,0.259579098865215);

\path [draw=color0, thick]
(axis cs:0.75,0.216190915033747)
--(axis cs:0.75,0.237009084966253);

\path [draw=color0, thick]
(axis cs:0.8,0.18790419239486)
--(axis cs:0.8,0.20809580760514);

\path [draw=color0, thick]
(axis cs:0.85,0.148479819285351)
--(axis cs:0.85,0.167386847381316);

\path [draw=color0, thick]
(axis cs:0.9,0.0995236446551199)
--(axis cs:0.9,0.115943022011547);

\path [draw=color0, thick]
(axis cs:0.95,0.0482368494668599)
--(axis cs:0.95,0.0601631505331401);

\path [draw=color0, thick]
(axis cs:1,0)
--(axis cs:1,0);

\path [draw=color1, thick]
(axis cs:0,0)
--(axis cs:0,0);

\path [draw=color1, thick]
(axis cs:0.05,0.0550439922144917)
--(axis cs:0.05,0.0589214214539847);

\path [draw=color1, thick]
(axis cs:0.1,0.118389877874204)
--(axis cs:0.1,0.123347915767894);

\path [draw=color1, thick]
(axis cs:0.15,0.177010651087443)
--(axis cs:0.15,0.184013351853416);

\path [draw=color1, thick]
(axis cs:0.2,0.24279885134393)
--(axis cs:0.2,0.252368074315057);

\path [draw=color1, thick]
(axis cs:0.25,0.317051682747616)
--(axis cs:0.25,0.329288704478111);

\path [draw=color1, thick]
(axis cs:0.3,0.399448492648235)
--(axis cs:0.3,0.413690959766065);

\path [draw=color1, thick]
(axis cs:0.35,0.475839886220209)
--(axis cs:0.35,0.49106531912331);

\path [draw=color1, thick]
(axis cs:0.4,0.555486801676371)
--(axis cs:0.4,0.570957096491749);

\path [draw=color1, thick]
(axis cs:0.45,0.616692296139339)
--(axis cs:0.45,0.631776069874404);

\path [draw=color1, thick]
(axis cs:0.5,0.659189154422225)
--(axis cs:0.5,0.673868846111247);

\path [draw=color1, thick]
(axis cs:0.55,0.686599425481465)
--(axis cs:0.55,0.700984355180309);

\path [draw=color1, thick]
(axis cs:0.6,0.676901113166158)
--(axis cs:0.6,0.692672202150752);

\path [draw=color1, thick]
(axis cs:0.65,0.632801567803593)
--(axis cs:0.65,0.649899431917042);

\path [draw=color1, thick]
(axis cs:0.7,0.539902318106556)
--(axis cs:0.7,0.557652776672044);

\path [draw=color1, thick]
(axis cs:0.75,0.435471604261861)
--(axis cs:0.75,0.452343041604836);

\path [draw=color1, thick]
(axis cs:0.8,0.32152775958831)
--(axis cs:0.8,0.335931754713917);

\path [draw=color1, thick]
(axis cs:0.85,0.215410648459734)
--(axis cs:0.85,0.225762719270441);

\path [draw=color1, thick]
(axis cs:0.9,0.123507578132628)
--(axis cs:0.9,0.130542240944817);

\path [draw=color1, thick]
(axis cs:0.95,0.0524004065784433)
--(axis cs:0.95,0.0564326067638824);

\path [draw=color1, thick]
(axis cs:1,0)
--(axis cs:1,0);

\addplot [thick, color0]
table {%
0 0
0.05 0.0737944444444444
0.1 0.1179
0.15 0.15675
0.2 0.197683333333333
0.25 0.230183333333333
0.3 0.2476
0.35 0.264
0.4 0.2866
0.45 0.2878
0.5 0.288
0.55 0.2984
0.6 0.2794
0.65 0.2712
0.7 0.249
0.75 0.2266
0.8 0.198
0.85 0.157933333333333
0.9 0.107733333333333
0.95 0.0542
1 0
};
\addplot [thick, color1]
table {%
0 0
0.05 0.0569827068342382
0.1 0.120868896821049
0.15 0.180512001470429
0.2 0.247583462829493
0.25 0.323170193612863
0.3 0.40656972620715
0.35 0.483452602671759
0.4 0.56322194908406
0.45 0.624234183006871
0.5 0.666529000266736
0.55 0.693791890330887
0.6 0.684786657658455
0.65 0.641350499860318
0.7 0.5487775473893
0.75 0.443907322933349
0.8 0.328729757151113
0.85 0.220586683865087
0.9 0.127024909538722
0.95 0.0544165066711628
1 0
};
\end{axis}

\end{tikzpicture}
  \caption{Variability of scheduled allocations.}
  \label{fig:variability}
 \end{subfigure}
 \hspace*{\fill}%
 \begin{subfigure}[b]{0.45\textwidth}
  \centering
\begin{tikzpicture}

\definecolor{color0}{rgb}{0.1725,0.4824,0.7137}
\definecolor{color1}{rgb}{0.9922,0.6824,0.3804}

\pgfplotsset{every tick label/.append style={font=\scriptsize}}
\tikzstyle{every node}=[font=\scriptsize]

\begin{axis}[
width=\fwidth,
height=\fheight,
at={(0,0)},
legend cell align={left},
legend style={font=\scriptsize, at={(0.03,0.03)}, anchor=south west, draw=white!80!black,
              /tikz/every even column/.append style={column sep=1em}},
tick align=outside,
tick pos=left,
x grid style={white!69.01960784313725!black},
xlabel={$P(C_1)$},
xmajorgrids,
xmin=-0.05, xmax=1.05,
xtick style={color=black},
y grid style={white!69.01960784313725!black},
ylabel={BI Occupancy Ratio},
ymajorgrids,
ymin=0.90, ymax=1.00220756063263,
ytick style={color=black},
xlabel style={font=\scriptsize\color{white!15!black}},
ylabel style={font=\scriptsize\color{white!15!black}}
]
\path [draw=color0, thick]
(axis cs:0,0.9996)
--(axis cs:0,0.9996);

\path [draw=color0, thick]
(axis cs:0.05,0.974467153565073)
--(axis cs:0.05,0.980823613101594);

\path [draw=color0, thick]
(axis cs:0.1,0.989352756319833)
--(axis cs:0.1,0.992886517013501);

\path [draw=color0, thick]
(axis cs:0.15,0.997324543322486)
--(axis cs:0.15,0.998900643344181);

\path [draw=color0, thick]
(axis cs:0.2,0.999075304511529)
--(axis cs:0.2,0.999632015488471);

\path [draw=color0, thick]
(axis cs:0.25,0.999311495216419)
--(axis cs:0.25,0.999647111450248);

\path [draw=color0, thick]
(axis cs:0.3,0.999601456285234)
--(axis cs:0.3,0.999608130381433);

\path [draw=color0, thick]
(axis cs:0.35,0.999605446689873)
--(axis cs:0.35,0.999612613310127);

\path [draw=color0, thick]
(axis cs:0.4,0.999610908686619)
--(axis cs:0.4,0.999618704646715);

\path [draw=color0, thick]
(axis cs:0.45,0.999621594602378)
--(axis cs:0.45,0.999629858730956);

\path [draw=color0, thick]
(axis cs:0.5,0.999631926606049)
--(axis cs:0.5,0.999640586727284);

\path [draw=color0, thick]
(axis cs:0.55,0.999639637146923)
--(axis cs:0.55,0.999648389519744);

\path [draw=color0, thick]
(axis cs:0.6,0.999652020522995)
--(axis cs:0.6,0.999661132810339);

\path [draw=color0, thick]
(axis cs:0.65,0.999662677526451)
--(axis cs:0.65,0.999671929140216);

\path [draw=color0, thick]
(axis cs:0.7,0.99967865532781)
--(axis cs:0.7,0.999687938005523);

\path [draw=color0, thick]
(axis cs:0.75,0.999690844945259)
--(axis cs:0.75,0.999699981721408);

\path [draw=color0, thick]
(axis cs:0.8,0.999703338056482)
--(axis cs:0.8,0.999712355276851);

\path [draw=color0, thick]
(axis cs:0.85,0.999524415973626)
--(axis cs:0.85,0.999733304026374);

\path [draw=color0, thick]
(axis cs:0.9,0.998161870670637)
--(axis cs:0.9,0.999306989329363);

\path [draw=color0, thick]
(axis cs:0.95,0.99503778178893)
--(axis cs:0.95,0.997178064877737);

\path [draw=color0, thick]
(axis cs:1,0.99979)
--(axis cs:1,0.99979);

\path [draw=color1, thick]
(axis cs:0,0.9995501)
--(axis cs:0,0.9995501);

\path [draw=color1, thick]
(axis cs:0.05,0.961628778850306)
--(axis cs:0.05,0.964254707816361);

\path [draw=color1, thick]
(axis cs:0.1,0.96808257554381)
--(axis cs:0.1,0.970059131122858);

\path [draw=color1, thick]
(axis cs:0.15,0.971449477889282)
--(axis cs:0.15,0.973054568777385);

\path [draw=color1, thick]
(axis cs:0.2,0.973260254932824)
--(axis cs:0.2,0.974566285067176);

\path [draw=color1, thick]
(axis cs:0.25,0.975021463570572)
--(axis cs:0.25,0.976146589762762);

\path [draw=color1, thick]
(axis cs:0.3,0.97584614195005)
--(axis cs:0.3,0.97686505804995);

\path [draw=color1, thick]
(axis cs:0.35,0.974922851929023)
--(axis cs:0.35,0.975977528070977);

\path [draw=color1, thick]
(axis cs:0.4,0.97508722464563)
--(axis cs:0.4,0.976139982021037);

\path [draw=color1, thick]
(axis cs:0.45,0.974214182818802)
--(axis cs:0.45,0.975310437181198);

\path [draw=color1, thick]
(axis cs:0.5,0.972717338025726)
--(axis cs:0.5,0.973902495307607);

\path [draw=color1, thick]
(axis cs:0.55,0.970536136361753)
--(axis cs:0.55,0.971899863638247);

\path [draw=color1, thick]
(axis cs:0.6,0.967506607754138)
--(axis cs:0.6,0.969081358912529);

\path [draw=color1, thick]
(axis cs:0.65,0.962252209287214)
--(axis cs:0.65,0.964269417379453);

\path [draw=color1, thick]
(axis cs:0.7,0.957624665661545)
--(axis cs:0.7,0.960183754338455);

\path [draw=color1, thick]
(axis cs:0.75,0.951475319382658)
--(axis cs:0.75,0.954458867284009);

\path [draw=color1, thick]
(axis cs:0.8,0.949671527132219)
--(axis cs:0.8,0.953037979534448);

\path [draw=color1, thick]
(axis cs:0.85,0.952924465558941)
--(axis cs:0.85,0.956037721107726);

\path [draw=color1, thick]
(axis cs:0.9,0.959803663090942)
--(axis cs:0.9,0.962437063575725);

\path [draw=color1, thick]
(axis cs:0.95,0.969479436276343)
--(axis cs:0.95,0.971850330390324);

\path [draw=color1, thick]
(axis cs:1,0.9997601)
--(axis cs:1,0.9997601);

\addplot [thick, color0]
table {%
0 0.9996
0.05 0.977645383333334
0.1 0.991119636666667
0.15 0.998112593333333
0.2 0.99935366
0.25 0.999479303333333
0.3 0.999604793333333
0.35 0.99960903
0.4 0.999614806666667
0.45 0.999625726666667
0.5 0.999636256666667
0.55 0.999644013333333
0.6 0.999656576666667
0.65 0.999667303333333
0.7 0.999683296666667
0.75 0.999695413333333
0.8 0.999707846666667
0.85 0.99962886
0.9 0.99873443
0.95 0.996107923333333
1 0.99979
};
\addplot [thick, color1]
table {%
0 0.9995501
0.05 0.962941743333333
0.1 0.969070853333334
0.15 0.972252023333333
0.2 0.97391327
0.25 0.975584026666667
0.3 0.9763556
0.35 0.97545019
0.4 0.975613603333333
0.45 0.97476231
0.5 0.973309916666667
0.55 0.971218
0.6 0.968293983333333
0.65 0.963260813333333
0.7 0.95890421
0.75 0.952967093333333
0.8 0.951354753333333
0.85 0.954481093333333
0.9 0.961120363333333
0.95 0.970664883333333
1 0.9997601
};
\end{axis}

\end{tikzpicture}
  \caption{BI occupancy ratio.}
  \label{fig:occupancy}
 \end{subfigure}
 \hspace*{\fill}%

 \caption{Results for Scenario 2.
  }
 \label{fig:results_scenario_2}
\end{figure*}

The first metric is the acceptance rate (\cref{fig:accrate_rho_bound}), defined as the ratio between the number of accepted allocation requests and the maximum number of acceptable requests.
To compute this achievable upper bound, since all allocations share the same parameters we ignore the strict periodicity assumption and calculate how many allocations with minimum duration $\Tblk{}=\Tmin{}$ can fit in a period $\Tp{}$, which is equal to $N_{\rm max}(\rho) = \left\lfloor \frac{\Tp{}}{\Tmin{}(\rho)} \right\rfloor$ and shown as a red, dashed line.
The acceptance rates can thus be normalized in the interval $[0,1]$, where $1$ means that the scheduler reaches the peak acceptance rate.
Since as $\rho \rightarrow 0$, $\Tmin{} \rightarrow 0$ and thus $N_{\rm max} (\rho) \rightarrow \infty$, we consider at most 100 allocations.

As expected, the \textit{simple scheduler} suffers from a lower acceptance rate than the \textit{max-min fair} one, even though, starting from $\rho=0.5$, the two algorithms tend to behave similarly.
In fact, more rigid allocations do not give enough flexibility to the \textit{max-min fair scheduler} to perform its optimization, thus yielding similar performance to the much \textit{simple scheduler}.

The second metric is \textit{Jain's Fairness Index} (\cref{fig:jain_rho}), defined as:
\begin{equation}
 \mathcal{J} = \frac{\left( \sum_n x_n \right)^2}{n \cdot \sum_n (x_n)^2},
\end{equation}
where only accepted allocations are counted and $x_n$ can either be the block duration $\Tblk{n}$ or the block duration ratio $r_n$.
If the $\{ x_n \}$ are all equal, then $\mathcal{J}(x)=1$.
On the other hand, the more unequal the values of $\{ x_n \}$, the closer the metric to its minimum $\mathcal{J}(x) = \frac{1}{N}$.

Based on the results plotted in \cref{fig:jain_rho}, both algorithms behave fairly with respect to the accepted allocations.

The third metric, shown in \cref{fig:tblk_rho}, offers a different perspective considering the average normalized block duration $\overline{\Tblk{}}/\Tmax{}(\rho)$.
As expected, the \textit{simple scheduler} shows an oscillating trend, due to the discrete behavior of the allocations.
In fact, the last scheduled allocation will only reduce \Tblk{} down to $\Tmin{}(\rho)$, thus, if the portion of DTI left by the previous allocations is less than $\Tmin{}(\rho)$, no additional allocations can be fitted.
On the other hand, the \textit{max-min fair scheduler} will try to reduce all allocations up to their minimum duration in order to avoid rejecting new ones, granting more accepted allocations at the cost of an overall lower block duration.

Finally, the two most discriminating metrics, namely the average normalized block duration and the acceptance rate, are plotted against each other in \cref{fig:tblk_accrate}.
In general, the \textit{simple scheduler} tends to favor higher average block duration for a lower acceptance rate, while the \textit{max-min fair scheduler} tends to favor acceptance rate at the cost of a lower average block duration, as expected.
Both algorithms are able to ensure high fairness to the accepted allocations, generally well above 0.85.

Similar behaviors were also observed for load factors $\lambda \in \{0.025, 0.4\}$, not shown here.
As expected, higher loads tend to have more pronounced variability in both the average block duration and the fairness granted to the accepted allocations.
Curiously, regardless of the load factor, for values of the interval ratio larger than $\rho\approx 0.5$, the two algorithms tend to have very similar performance due to the more rigid allocation requests that do not allow the \textit{max-min fair scheduler} to exploit its agility.

\ml{what about $\Tp{}>\TBI$? does anything change?} \md{if I recall correctly, we obtained similar results and so we avoid further studies on this case}

\textit{Scenario~2} allows us to analyze the impact of allocations with different periodicities on the overall network performance, as a function of the probability $P(C_1)$ that a request $C_1$ is offered to the system.

Since allocations with different periods coexist, it is mandatory to decide how many allocations should be offered to the schedulers, as this will affect how the acceptance rate is normalized.
We define the \textit{minimum occupancy} as the minimum BI occupancy ratio of the allocation of category $C_i$, namely $O_{\rm min}^i = \Tmin{i}/\Tp{i}$.
Allocations are offered to the schedulers as long as the cumulative minimum offered occupancy does not exceed the value of 1.

In \cref{fig:conditional_c1,fig:conditional_c2} we show the biasing effect of the first accepted allocation on the proposed schedulers.
Clearly, the \textit{simple scheduler} suffers a strong and symmetric effect, meaning that once the first allocation is scheduled with maximum duration, it will be harder for subsequent allocations with a different period to fit the constrained BI, making the scheduler favor allocations with the same period.
On the other hand, it is significantly harder to interpret the behavior of the more complex \textit{max-min fair scheduler}.
From further results, not shown here for lack of space, it is possible to notice that allocations with a lower average BI occupancy are favored, with a slight preference towards those with lower values of \Tp{} and \Tmin{}.
As also shown here, in fact, allocations with lower values of \Tp{}, such as $C_2$ with respect to $C_1$, tend to fragment the BI more, making it harder to then fit allocations with different periodicity and higher \Tmin{}.

To further confirm this biasing behavior, we show the variability $\nu$ among the scheduled allocations, defined as
\begin{equation}
  \nu_{1,2}=\frac{\min\{| C_1 |, | C_2 |\}}{\max\{| C_1 |, | C_2 |\}},
\end{equation}
where $| C_i |$ represents the number of accepted allocations of category $C_i$.
This metric takes values in $[0,1]$, where a value of 0 means that only allocations of a single type have been accepted, while a value of 1 means that the same number of allocations of both categories have been accepted.
Results are shown in \cref{fig:variability}, which confirms that the \textit{simple scheduler} favors a more homogeneous BI allocation, while the \textit{max-min fair scheduler} shows once again more flexibility, being able to accommodate fairly requests from different classes, as shown by the higher variability of the accepted allocations.

Finally, we studied how efficiently the two algorithms are able to use the radio resources by measuring the BI occupancy ratio, i.e., the ratio between scheduled and unscheduled air time.
It can be noticed that, while the \textit{simple scheduler} accepts fewer requests, it is able to use almost all available resources.
This is due to the fact that the scheduler tends to accept homogeneous allocations, allowing them to be packed more efficiently in the BI.
On the other hand, the \textit{max-min fair scheduler} successfully fits multiple allocations of both types, but the constraints on the periodicity and the minimum block duration $\Tmin{}$ prevent it from fully utilizing the whole \gls{bi} when a mixture of the two types of sources is presented.
Nonetheless, it ensures very high occupancy ratios, always above 95\% for the shown example.


\section{Conclusions} 
\label{sec:conclusions}

In this paper, we presented a framework for periodic scheduling in \gls{wigig}-compatible devices.
We proposed two heuristic algorithms, \textit{simple} and \textit{max-min fair schedulers}, and accurately described their inner workings.
Finally, we assessed their performance in two different scenarios, showing that the \textit{max-min fair scheduler} tends to trade resource availability for a much higher acceptance rate, contrary to the \textit{simple scheduler}'s behavior, while both schedulers obtained a high Jain's fairness index for the accepted allocations.

Even working in a simplified settings without considering further sources of complexity from other parts of the communication stack, it was possible to notice that both the design and the evaluation of \gls{wigig}-specific scheduling algorithms for periodic sources is highly non-trivial and can show surprising results.
In fact, while the formalization of the problem is straightforward, scheduling algorithms often have to deal with many hard-to-predict edge cases, which greatly increases the difficulty of designing a general algorithm.

Future works will focus on multiple objectives.
A first objective is to implement these algorithms in a full-stack simulator, allowing the study of APP-layer performance metrics for a range of possible applications.
A second objective is to extend the framework by relaxing some of the assumptions made in \cref{sec:framework_description} and test the impact of each one of them on the overall performance of a \gls{wigig} system.
A third objective is to extend the study for scheduling multiple allocations at once, a possibility given by \gls{mimo} techniques.
To further achieve a realistic evaluation, the impact of accurate wireless channel simulations, physical layer design, as well as user mobility will also be studied.


\bibliographystyle{IEEEtran}
\bibliography{bibl.bib}

\end{document}